# Analytic binary alloy volume-concentration relations and the deviation from Zen`s law

A. Landa, J.E. Klepeis, R.E. Rudd, K.J. Caspersen, and D.A. Young

Lawrence Livermore National Laboratory, Livermore, CA 94551 USA

**Abstract**

Alloys expand or contract as concentrations change, and the resulting relationship between atomic volume and alloy content is an important property of the solid. While a well-known approximation posits that the atomic volume varies linearly with concentration (Zen`s law), the actual variation is more complicated. Here we use an *apparent* size of the solute (solvent) atom and the elasticity to derive explicit analytical expressions for the atomic volume of binary solid alloys. Two approximations, *continuum* and *terminal*, are proposed. Deviations from Zen`s law are studied for 22 binary alloy systems.

## 1. Introduction

According to the Hume-Rothery rules [1], a) extensive substitutional solid solution occurs only if the relative difference between the atomic diameters (radii) of the two species is less than 15%.; b) for appreciable solid solubility, the crystal structures of the two elements must be identical; c) a metal will dissolve a metal of higher valency to a greater extent than one of lower valency; d) an electronegativity difference close to zero gives maximum solubility. The more electropositive one element and the more electronegative the other, the greater the likelihood is that they will form an intermetallic compound instead of a substitutional solid solution.

According to Vegard's law [2], unit cell parameters should vary linearly with composition for a continuous substitutional solid solution in which the atoms or ions that substitute for each other are randomly distributed. For ideal solutions, with excess energies and volumes equal to zero, the atomic volumes ($\Omega$) of disordered alloys vary linearly with composition (Zen`s law, [3]): $\Omega(x) = \Omega_A x + \Omega_B(1-x)$, where $x$ is atomic composition of the element $A$. One should notice that even for an ideal solution (for which Zen`s law is valid), a deviation from Vegard's law can occur. Indeed, for a cubic structure, the lattice parameter, $a$, is related to the cell volume, $V$, by the relation, $a = V^{1/3}$, and linear variation of the cell volume should not imply linear variation of the lattice parameter.

Models exist to predict the deviation from Zen`s law; however, they are not sufficiently reliable. Even the probability to predict the sign of the deviation from Zen`s law does not exceed 60 %. Hume-Rothery and Raynor [4] found a significant negative deviation from Zen`s law for Cd-Mg solid solution. Massalski and King [5] found the numerous intermetallic phases in Cu-Zn system show a negative deviation from Zen`s law. Indeed, the negative deviations from Zen`s law is observed for most of the ordered compounds. The excess volume of the alloy, $\Delta\Omega = \Omega - \Omega_{Zen`s}$, usually called "superstructure contraction", plays an important role in stabilization of intermediate structures due to the free energy gain ($\Omega_{Zen`s}$ is the atomic volume determined by Zen`s law).

Kozlov *et al.* [6] introduced several parameters that define geometry of binary alloys. In addition to the dimensional size factor, $\delta = R_A/R_B$, where $R_A$ and $R_B$ are the atomic radii of components A and B, respectively, Lawes-Parthé space filling factor, $\psi$, and superstructure contraction, $\Delta\Omega/\Omega$, have been introduced. The space filling factor of a binary alloys is defined as $\psi = [\Omega_A x + \Omega_B(1-x)]/\Omega(x)$. By studying numerous binary alloys with AB and A$_3$B stoichiometry, Kozlov *et al.* [6] found that among AB stoichiometry structures, B2, B19, and L1$_0$ structures with *Pm3m*, *Pmcm,* and *C4/mmm* Space Groups, respectively, show a negative superstructure contraction, $\Delta\Omega/\Omega$, equal to - 0.100, - 0.038, - 0.045, respectively, and only $L1_1$ structure, the Space Group R-3m, show a very small but positive superstructure contraction equal to + 0.003. Among $A_3B$ stoichiometry structures, $L1_2$ (*Pm3m*), $D0_{19}$ (P$6_3$/mmc), $D0_{23}$ *(4I/mmm)*, and $D0_{24}$ *(P6$_3$/mmc)* structures show a negative superstructure contraction equal to - 0.153, - 0.053, - 0.105, and - 0.144, respectively. Two $A_3B$ structures, $D0_{22}$ *(4I/mmm)*, and $A_{15}$ *(I/mmm)* show a positive superstructure contraction equal to + 0.011 and + 0.054, respectively. The existence of the stable intermetallic compounds with a positive superstructure contraction is explained in terms of the space filling factor, $\psi$.

Kozlov *et al.* [7] have constructed the histograms of distribution of the B2 and $L1_2$ superstructures in Ni-Al system as a function of the space filling factor, $\psi$, and superstructure contraction, $\Delta\Omega/\Omega$. It was found that the space filling factor and enthalpy of phase formation increase simultaneously in Ni-Al intermetallic compounds. The space filling factor and superstructure contraction intertwine: the higher the space filling factor, the higher the superstructure contraction. It was shown that the rate of change in superstructural contraction determines, in many respects, the enthalpy of phase formation: the higher space filling factor, the higher the enthalpy of phase formation. The highest the space filling factor, superstructure contraction, and the enthalpy of formation are observed for stoichiometric NiAl (B2) structure: 0.785, - 0.140, and - 59 kJ/mol, respectively.

Klopotov *et al*. [8] have addressed the main crystallogeometrical parameters of compounds in the Ni–Ti system. It has been found out that the space filling factor, superstructural contraction, and enthalpy of phase formation increase simultaneously. The highest space filling factor, superstructure contraction, melting temperature and the enthalpy of formation are observed for stoichiometric $Ni_3Ti$ ($D0_{24}$) structure: 0.80, -0.086, 1400 ºC, and - 38 kJ/mole, respectively.

Potekaev *et al*. [9, 10] have established the explicit correlation between the type of evolution of the binary phase diagrams based on elements of VIIIA and IB groups of the Mendeleyev's Periodic Table and the nature (positive/negative) of the superstructure contraction.

The superstructure contraction, which reflects the deviation from Zen`s law, is a very important parameter of the binary alloy crystal lattice. In the next Section we discuss a theoretical basis, based on Lubarda's [11] elastic inclusion model, for analytical determination of the atomic volume of solid solutions.

## 2. Theoretical Background

Lubarda [11] derived an expression for the *effective* lattice parameter of binary solid solutions by using an elasticity inclusion model, in conjunction with an *apparent* size of the *solute* atom when resolved in the *solvent* matrix. Assuming the $R_1$ is the Wigner-Seitz radius of the solvent material, $\Omega_1 = \frac{4\pi R_1^3}{3}$, the volume increase produced by insertion of the solute atom to the solvent is expressed as

$$\Delta \Omega = 4\pi R_1^3 \gamma_1 C, \tag{1}$$

where the misfit coefficient, $C$, is given by the expression

$$C = \frac{(R_2^* - R_1)}{R_1 \gamma_2} \tag{2}$$

and $R_2^*$ is the *apparent* Wigner-Seitz radius of the solute material, which is introduced to approximately account for the electronic interactions between the outermost quantum shells of the solute and solvent atoms. The parameters $\gamma_1$ and $\gamma_2$ are defined by

$$\gamma_1 = 1 + \frac{4\mu_1}{3K_1}, \quad \gamma_2 = 1 + \frac{4\mu_1}{3K_2}, \tag{3}$$

where $\mu_{1(2)}$ and $K_{1(2)}$ are the shear and bulk moduli of the solvent (solute). If $x$ is the atomic concentration of the solute, the total volume increase produced by insertion of the $N_2 = xN_1$ solute atoms ($N_1$ is the total number of the solvent atoms) is

$$\Delta V = 4\pi R_1^3 x N_1 \gamma_1 C \tag{4}$$

and the total volume is

$$V = V_1 + 4\pi R_1^3 x N_1 \gamma_1 C, \tag{5}$$

where $V_1$ is the volume of the solvent. The atomic volume of the system with $N_2 = xN_1$ solute atoms will be

$$\Omega = \Omega_1 + 4\pi R_1^3 x \gamma_1 C. \tag{6}$$

The *apparent* radius of the solute atom, $R_2^*$, is estimated using one piece of experimental information about the solid solution, i.e. the initial slope of the lattice spacing vs. composition curve, $\left(\frac{da}{dx}\right)_{x=0}$. King [12] introduced the volume size factor,

$$\omega_2 = \frac{1}{\Omega_1}\left(\frac{d\Omega}{dx}\right)_{x=0} = \frac{3}{a_1}\left(\frac{da}{dx}\right)_{x=0}. \tag{7}$$

By differentiating equation (6), $\Omega = \Omega_1 + 4\pi R_1^3 x \gamma_1 C$, one gets

$$\left(\frac{d\Omega}{dx}\right)_{x=0} = 4\pi R_1^3 \gamma_1 C. \tag{8}$$

Thus, combining expressions (7) and (8), one gets $\omega_2 \Omega_1 = \left(\frac{d\Omega}{dx}\right)_{x=0} = 4\pi R_1^3 \gamma_1 C = 4\pi R_1^3 \gamma_1 \frac{(R_2^* - R_1)}{R_1 \gamma_2}$ or $\omega_2 \Omega_1 = 3 \frac{4\pi R_1^3}{3} \gamma_1 \frac{(R_2^* - R_1)}{R_1 \gamma_2} = 3\Omega_1 \gamma_1 \frac{(R_2^* - R_1)}{R_1 \gamma_2}$ and $\omega_2 = 3\gamma_1 \frac{(R_2^* - R_1)}{R_1 \gamma_2}$. Thus, the *apparent* Wigner-Seitz radius of the solute atom can be calculated from

$$R_2^* = R_1 \left(1 + \frac{\gamma_2}{3\gamma_1} \omega_2\right) \tag{9}$$

and an analogous expression is derived for $R_1^* = R_2 \left(1 + \frac{\gamma_1}{3\gamma_2} \omega_1\right)$. King [12], using the high-precision lattice parameter data available in the literature, presented numerical values of $\omega_i$ for 469 metallic solid solutions. The values of the *experimental* Wigner-Seitz radii, $R_1$, $R_2$, measures at room temperatures, the volumes size factors, $\omega_1$ and $\omega_2$, and the *apparent* radii, $R_1^*$ and $R_2^*$, were listed in Tables 1 and 3 of Ref. [11].

According to Lubarda [11] and Eq. (6), the *effective* lattice parameter, $a$, of a binary alloy is

$$a = [a_1^3 + 4\pi R_1^3 x \frac{\kappa_1}{\vartheta_1} \gamma\, C]^{1/3}, \tag{10}$$

where

$$\gamma = 1 + \frac{4\mu_1}{3K}, \tag{11}$$

where $K$ is the *effective* bulk modulus and $\kappa$ is the number of atoms per unit cell used to define the atomic volume $\Omega = \frac{\vartheta a^3}{\kappa}$, where $a_1$ denotes the lattice parameter of the solvent, and $\kappa$ is equal to 1 for the simple cubic (SC), 2 for the body-centered cubic (BCC), 4 for the face-centered cubic (FCC), 6 for the hexagonal close-packed (HCP), and 8 for the diamond-cubic lattice (DCL) structure. The parameter $\vartheta = 1$ for cubic lattices, and $\vartheta = 3\sqrt{3}\frac{c}{2a}$ for the ideal hexagonal close-packed lattice. Lubarda [11] suggested performing calculations of the *effective* lattice parameter of a binary alloy in two stages based on micromechanics. First, assume that $a_1$ is the lattice constant of the solvent and $a_2$ is the lattice constant of the solute. In this case, the *effective* lattice parameter becomes:

$$a = [a_1^3 + 3x\frac{\gamma}{\gamma_2} a_1^2 (\zeta a_2 - a_1)]^{1/3}, \tag{12}$$

where

$$\frac{\gamma}{\gamma_2} = \frac{1 + 4\mu_1/3K}{1 + 4\mu_1/3K_2} \tag{13}$$

and

$$\zeta = \frac{R_2^*}{R_2} \sqrt[3]{\frac{\vartheta_2 \kappa_1}{\vartheta_1 \kappa_2}}. \tag{14}$$

The terminal solid solution at the other end of the phase diagram can be treated by reversing the role of two materials ($a_2$ is the lattice constant of the solvent and $a_1$ is the lattice constant of the solute). Thus, Eq. (12) is replaced with

$$a = [a_2^3 + 4\pi R_2^3 (1-x) \frac{\kappa_2}{\vartheta_2} \gamma\, C]^{1/3}, \tag{15}$$

where

$$\gamma = 1 + \frac{4\mu_2}{3K} \tag{16}$$

and

$$C = \frac{(R_1^* - R_2)}{R_2 \gamma_1}. \tag{17}$$

In that case, Eq. (15) becomes

$$a = [a_2^3 + 3(1-x) \frac{\gamma}{\gamma_1} a_2^2 (\zeta a_1 - a_2)]^{1/3}, \tag{18}$$

where

$$\frac{\gamma}{\gamma_1} = \frac{1 + 4\mu_2/3K}{1 + 4\mu_2/3K_1} \tag{19}$$

and

$$\zeta = \frac{R_1^*}{R_1} \sqrt[3]{\frac{\vartheta_1 \kappa_2}{\vartheta_2 \kappa_1}}. \tag{20}$$

To use Eqs. (12, 18) one must calculate the *effective* shear ($\mu$) and bulk ($K$) moduli. Lubarda [11] used Hill's self-consistent method, as presented by Nemat-Nasser and Hori, [13] that gives the following system of equations for the *effective* shear and bulk moduli:

$$\frac{1-x}{1+4\mu/3K_1} + \frac{x}{1+4\mu/3K_2} - 5\left(\frac{1-x}{1-\frac{\mu}{\mu_2}} + \frac{x}{1-\frac{\mu}{\mu_1}}\right) + 2 = 0, \tag{21}$$

$$K(x) = \left(\frac{1-x}{k_1 + 4\mu/3} + \frac{x}{k_2 + 4\mu/3}\right)^{-1} - \frac{4}{3}, \tag{22}$$

and the atomic volume for the alloys can be calculated as

$$\Omega = \Omega_1 + 4\pi R_1^3 x \gamma C, \tag{23}$$

where $\gamma = 1 + \frac{4\mu_1}{3K}$ is written by Eq. (11).

However, if Eq. (22) has a reasonable solution in both terminal cases, $x = 0$, $K = K_1$ and $x = 1$, $K = K_2$, Eq. (21) has negative solutions in both terminal cases:

$$x = 0, \mu = \frac{-3(3K_1 + 4\mu_1) \pm \sqrt{9(3K_1 + 4\mu_1)^2 - 192 K_1 \mu_1}}{16}; \quad x = 1, \mu = \frac{-3(3K_2 + 4\mu_2) \pm \sqrt{9(3K_2 + 4\mu_2)^2 - 192 K_2 \mu_2}}{16}.$$

In the present calculations we use the Voigt-Reuss-Hill approximation [14] to calculate the *effective* shear modulus, $\mu_{VRH}(x)$:

$$\mu_V(x) = (1-x)\mu_1 + x\mu_2$$

$$\mu_R(x) = \left[\frac{(1-x)}{\mu_1} + \frac{x}{\mu_2}\right]^{-1} \qquad (24)$$

$$\mu(x) = \mu_{VRH}(x) = \frac{1}{2}(\mu_V(x) + \mu_R(x))$$

Equations (22) and (23) are solved self-consistently. For a special case of the small atomic volume misfit, Lubarda [11] assumed:

$$\Omega_2 - \Omega_1 = 4\pi R_1^3 \gamma_2 C, \qquad (25)$$

which allows rewriting Eq. (23):

$$\Omega = \Omega_1 + 4\pi R_1^3 x\gamma C = \Omega_1 + 4\pi R_1^3 \gamma_2 Cx\left(\frac{\gamma}{\gamma_2}\right) = \Omega_1 + (\Omega_2 - \Omega_1)\frac{\gamma}{\gamma_2}x,$$

so

$$\Omega = \Omega_1 + (\Omega_2 - \Omega_1)\frac{\gamma}{\gamma_2}x. \qquad (26)$$

Lubarda wrote this expression in the slightly different form

$$\Omega = \left(1 - \frac{\gamma}{\gamma_2}x\right)\Omega_1 + \frac{\gamma}{\gamma_2}x\Omega_2, \qquad (27)$$

emphasizing that in the case of $\gamma = \left(1 + \frac{4\mu_1}{3K}\right) = \gamma_2 = \left(1 + \frac{4\mu_1}{3K_2}\right)$, which, according to Eqs. (3) and (11), means if $K=K_2$, Zen`s mixture rule of additive atomic volumes of the solute and solvent will be fulfilled.

In this paper several additional assumptions have been made to achieve a continuum solution for the alloy atomic volume within the whole composition range. As pointed out above, see Eqs. (10, 15), Lubarda assumed that calculations should be performed for two opposite terminal solid solutions located on the ends of the binary phase diagram. In that case the calculated lattice constants typically have a discontinuity in the middle, at the equiatomic composition. To avoid this problem, we redefine (symmetrize) the coefficient $\gamma_1$, Eq. (3), as well as assume that the *effective* coefficient, $\gamma$, defined by Eq. (11), should be recalculated at each alloy composition, $x$, and be expressed through the *effective* bulk, $K(x)$, and shear, $\mu(x)$, moduli calculated by Eqs. (22, 24):

$$\gamma_1 = 1 + \frac{4\mu_2}{3K_1},\ \gamma_2 = 1 + \frac{4\mu_1}{3K_2} \qquad (28)$$

$$\gamma(x) = 1 + \frac{4\mu(x)}{3K(x)}. \qquad (29)$$

In addition, we have rewritten Eq. (26) for the two terminal solid solutions:

$$\Omega^{(1)}(x) = \Omega_1 + (\Omega_2 - \Omega_1)\frac{\gamma(x)}{\gamma_2}x, \tag{30}$$

$$\Omega^{(2)}(x) = \Omega_2 + (\Omega_1 - \Omega_2)\frac{\gamma(x)}{\gamma_1}(1-x). \tag{31}$$

And the atomic volume of the alloy is defined as a function of composition, $x$,

$$\Omega(x) = \Omega^{(1)}(x) + x\left(\Omega^{(2)}(x) - \Omega^{(1)}(x)\right). \tag{32}$$

Lubarda [11] used the *apparent* Wigner-Seitz radius, Eq. (9), in calculations of the lattice parameters of the alloy. In our calculations we introduce two approximations for the atomic volumes of the alloy components which, in turn, are used as input parameters in Eq. (32).

1. *Continuum* approximation. In the case, where the field of the disordered solid solution spans throughout the whole composition range, we assume that the atomic volume of the solvent ($\Omega_1(x)$) changes linearly with composition from the *real* value, $\Omega_1$, $x = 0$, to its *apparent* value, $\Omega_1^*$, in the pure solute, $x = 1$.

$$\Omega_1(x) = \Omega_1(1-x) + \Omega_1^* x \tag{33}$$

Similarly, the atomic volume of the solute, $\Omega_2(x)$, changes linearly with composition, from its *apparent* value in the pure solvent, $\Omega_2^*$, $x = 0$, to the *real* value, $\Omega_2$, $x = 1$.

$$\Omega_2(x) = \Omega_2 x + \Omega_2^*(1-x) \tag{34}$$

2. *Terminal* approximation. In the case of limited mutual solubility of the alloy components, it is reasonable to consider the atomic volume of the solvent to be constant and equal to its *real* value, $\Omega_{1(2)}$,

$$\Omega_{1(2)}(x) = \Omega_{1(2)} \tag{35}$$

The atomic volume of the solute, $\Omega_{2(1)}$, undergoes a linear change with composition, $x$,

$$\Omega_{2(1)}(x) = \Omega_{2(1)}x + \Omega_{2(1)}^*(1-x). \tag{36}$$

The experimental (*real*) atomic volumes of selected elements at room temperature, together with the bulk and shear moduli are listed in Table 1. The atomic volumes correspond to the Wigner-Seitz radii reported in Table 1 of Ref. [11], and the elastic constants are the same as in Table 3 of Ref. [11]. In addition to the binary alloys studied in Ref. [11], we present data for Mg-Cd and Fe-Cr solid solutions. The volume size factors, $\omega_1$ and $\omega_2$, for the alloy systems under consideration are reproduced in Table 2. The values of the *real* and *apparent* atomic volumes for 22 alloys are listed in Table 3.

Table 1. The experimental atomic volumes and elastic constants for polycrystalline metals.

| Element | Ω (Å³) | K (GPa) | μ (GPa) |
| --- | --- | --- | --- |
| Mg | 23.2396 | 35.6 | 17.3 |
| Cd | 22.0210 | 46.8 | 19.1 |
| Al | 16.6036 | 72.6 | 26.0 |
| Si | 20.0182 | 97.6 | 66.2 |
| Ti | 17.6542 | 108.2 | 45.6 |
| V | 13.8256 | 157.9 | 46.7 |
| Cr | 12.0064 | 160.0 | 115.1 |
| Mn | 12.2199 | 98.0 | 39.0 |
| Fe | 11.7771 | 169.6 | 81.4 |
| Co | 11.0732 | 82.3 | 88.8 |
| Ni | 10.9415 | 183.0 | 80.0 |
| Cu | 11.8072 | 136.4 | 46.8 |
| Zn | 15.2123 | 69.6 | 41.9 |
| Ge | 22.6345 | 75.0 | 54.9 |
| Zr | 23.2790 | 94.0 | 30.0 |
| Nb | 17.8715 | 170.3 | 37.5 |
| Mo | 15.5834 | 261.3 | 125.5 |
| Ag | 17.0578 | 103.4 | 30.3 |
| Sn | 27.3255 | 58.2 | 18.4 |
| Ta | 18.0173 | 196.5 | 69.0 |
| W | 15.8566 | 311.0 | 160.6 |
| Au | 16.9618 | 170.7 | 27.5 |
| Pb | 30.3246 | 45.9 | 5.6 |

Table 2. The volume size factor data: $\omega_1$ is the volume size factor when the first element of the alloy system is the solute, and $\omega_2$ when the second element is the solute. Ref. [11, 12].

| Alloy | $\omega_1$ | $\omega_2$ |
|---|---|---|
| Al-Ag | − 0.0918 | + 0.0012 |
| Al-Cu | + 0.2000 | − 0.3780 |
| Al-Mg | − 0.3580 | + 0.4082 |
| Al-Mn | + 0.1620 | − 0.4681 |
| Al-Ti | − 0.2009 | − 0.1506 |
| Al-Zn | − 0.0625 | − 0.0574 |
| Cu-Ag | − 0.2775 | + 0.4352 |
| Cu-Au | − 0.2781 | + 0.4759 |
| Cu-Fe | + 0.1753 | + 0.0457 |
| Cu-Ni | + 0.0718 | − 0.0845 |
| Cu-Zn | − 0.5457 | + 0.1710 |
| Fe-Co | + 0.0524 | + 0.0154 |
| Fe-Cr | −0.0207 | + 0.0436 |
| Fe-V | − 0.1886 | + 0.1051 |
| Ag-Au | − 0.0064 | − 0.0178 |
| Ag-Mg | − 0.6342 | + 0.0713 |
| Mg-Cd | − 0.0160 | − 0.2108 |
| Si-Ge | − 0.2065 | + 0.0468 |
| Nb-Ta | − 0.0023 | − 0.0026 |
| Pb-Sn | + 0.2905 | − 0.0825 |
| Ti-Zr | − 0.2233 | + 0.3008 |
| Cr-W | − 0.2173 | + 0.3735 |

Table 3. The *real* (experimental) atomic volumes ($\Omega_1$ and $\Omega_2$) and *apparent* atomic volumes ($\Omega_1^*$ and $\Omega_2^*$) for the selected binary alloys.

| Alloy | $\Omega_1$ (Å³) | $\Omega_2$ (Å³) | $\Omega_1^*$ (Å³) | $\Omega_2^*$ (Å³) |
|---|---|---|---|---|
| Al-Ag | 16.6036 | 17.0578 | 15.3642 | 16.6193 |
| Al-Cu | 16.6036 | 11.8072 | 15.0820 | 11.8247 |
| Al-Mg | 16.6036 | 23.2396 | 17.2016 | 27.4032 |
| Al-Mn | 16.6036 | 12.2199 | 14.5684 | 10.4510 |
| Al-Ti | 16.6036 | 17.6542 | 13.8034 | 14.4677 |
| Al-Zn | 16.6036 | 15.2123 | 14.2961 | 15.6559 |
| Cu-Ag | 11.8072 | 17.0578 | 13.0147 | 18.4090 |
| Cu-Au | 11.8072 | 16.9618 | 12.4972 | 17.8913 |
| Cu-Fe | 11.8072 | 11.7771 | 14.1683 | 12.3236 |
| Cu-Ni | 11.8072 | 10.9415 | 11.8510 | 10.9128 |
| Cu-Zn | 11.8072 | 15.2123 | 9.5991 | 14.6320 |
| Fe-Co | 11.7771 | 11.0732 | 11.4817 | 12.0343 |
| Fe-Cr | 11.7771 | 12.0064 | 11.7664 | 12.3105 |
| Fe-V | 11.7771 | 13.8256 | 11.4252 | 13.0978 |
| Ag-Au | 17.0578 | 16.9618 | 16.8408 | 16.7868 |
| Ag-Mg | 17.0578 | 23.2396 | 13.9318 | 18.9916 |
| Mg-Cd | 23.2396 | 22.0210 | 21.6319 | 19.4024 |
| Si-Ge | 20.0182 | 22.6345 | 18.7414 | 21.1106 |
| Nb-Ta | 17.8715 | 18.0173 | 17.9742 | 17.8253 |
| Pb-Sn | 30.3246 | 27.3255 | 36.8260 | 27.9621 |
| Ti-Zr | 17.6542 | 23.2790 | 18.6324 | 23.8674 |
| Cr-W | 12.0064 | 15.8566 | 11.5458 | 15.7591 |

## 3. Results.

In this section we report calculations of the atomic volume as a function of concentration for 22 binary alloy systems chosen because of the availability of experimental data. All calculations and data are at room temperature. Both *continuum* and *terminal* approximations are applied.

### 3.1. Al-Ag.

According to the Refs. [11, 15], the maximum solubility of silver in aluminum is about 20 at.% at the eutectic temperature (567 °C), Fig. 1a. The lattice parameter of aluminum based solid solution remains practically unchanged up to 10 at % Ag. Maximum solubility of aluminum in silver is also about 20 at.% and occurs over a wider range of temperatures (450 °C - 610 °C) [11, 15], Fig. 1a. The calculated, in the *terminal* approximation, atomic volume of Al based solid solution is in good agreement with experimental data although the *continuum* approximation significantly underestimates the calculated atomic volume, Fig. 1b. For Ag based solution, both *continuum* and *terminal* approximations produce a significant negative deviation from Zen`s law which is in an accord with experimental observation; however, in this case, the atomic volume calculated in the *continuum* approximation almost exactly matches experimental results and the *terminal* approximation overestimates the atomic volume, Fig 1b. The significant negative deviation from Zen`s law for Ag based solid solution correlates with the negative heat of mixing observed in disordered Al-Ag alloys [38].

### 3.2. Al-Cu.

There is very limited solubility (~ 2.5 at.%, at 550 °C) of copper in aluminum [11, 16], Fig. 2a. Maximum solubility of aluminum in copper is about 20 at.% and occurs over a wide range of temperatures (360 °C - 567 °C) [11, 16], Fig 2a. Good agreement between the calculations (both *continuum* and *terminal* approximations) occurs at both ends of the concentration range, Fig. 2b. However contrary to results of calculation of the lattice parameter of the FCC Al-Cu solid solutions reported by Lubarda [11], which show a jump in the lattice parameter at the equiatomic composition, the *continuum* approximation eliminates this kind of jump in the atomic volume. As in the case of Al-Ag solid solutions, the negative deviation from Zen`s law for Cu based solid solution correlates with the negative heat of mixing observed in disordered Al-Cu alloys [38].

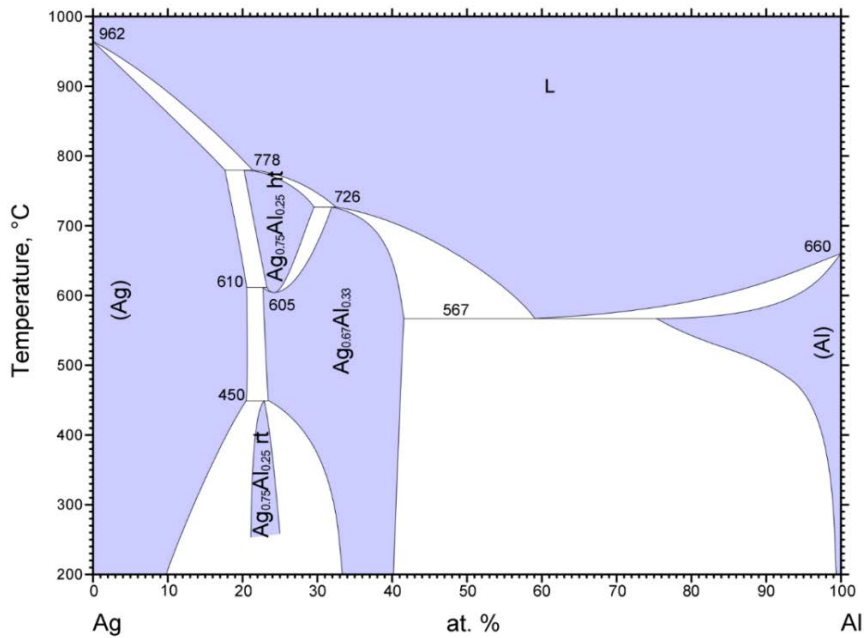

Fig 1a. Ag-Al phase diagram [15]. This plot is taken from ASM Alloy phase Diagram Database.

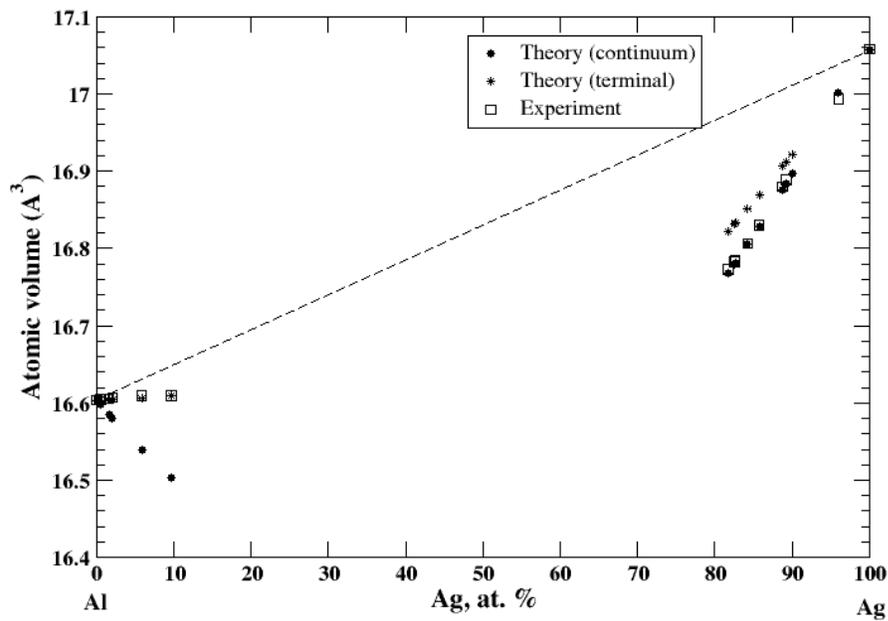

Fig 1b. Atomic volume vs. concentration for Al-Ag alloy system. The experimental data are from Ref. [37], pp. 261 and 351.

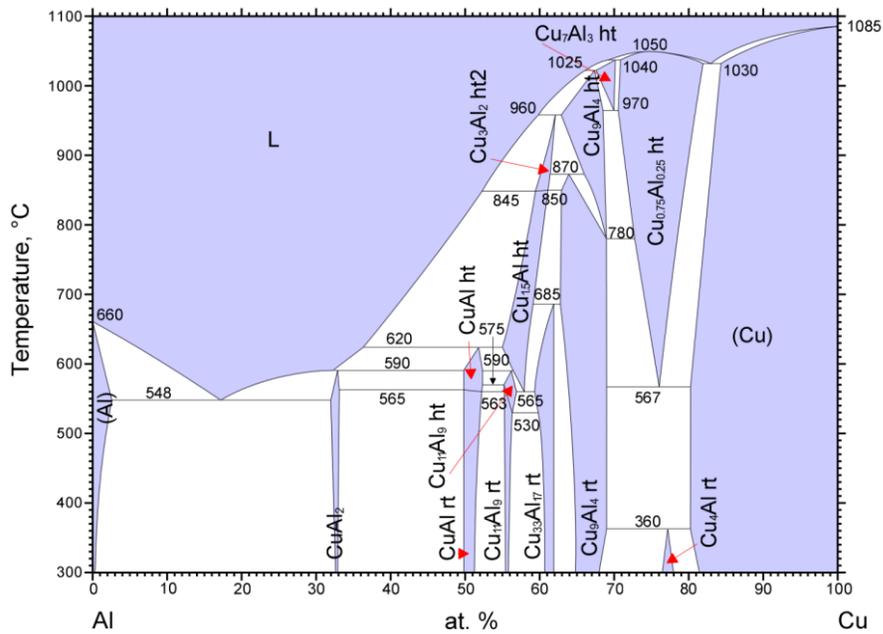

Fig 2a. Al-Cu phase diagram [16]. This plot is taken from ASM Alloy phase Diagram Database.

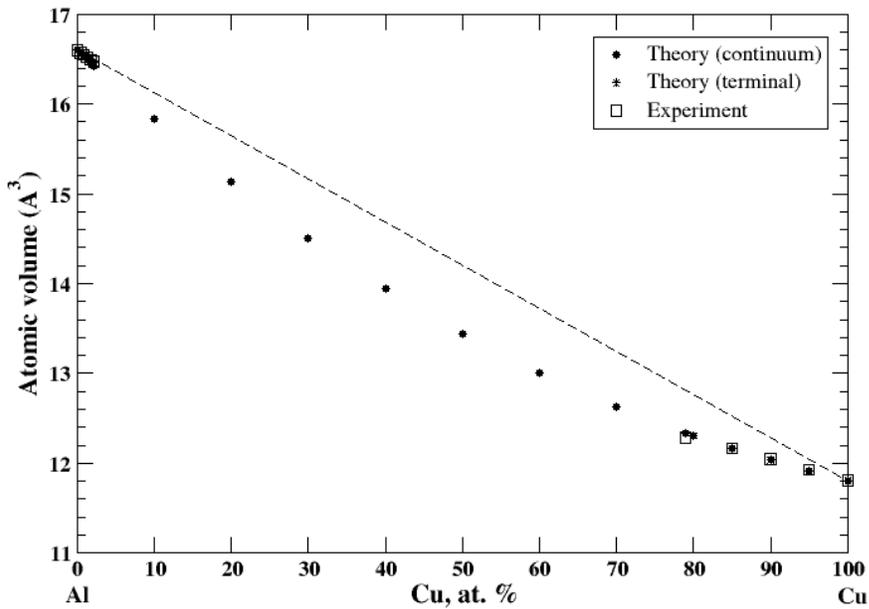

Fig 2b. Atomic volume vs. concentration for Al-Cu alloy system. The experimental data are from Ref. [37], pp. 328 and 331.

### 3.3. Al-Mg.

According to Refs. [11, 17], maximum solubility of magnesium in aluminum is about 18 at.% at 450 °C, and that of aluminum in magnesium is about 12 at.% at 437 °C, Fig 3a. According to Lubarda [11], the lattice spacing of Al based alloys increases by introduction of larger Mg atoms and the lattice spacing of Mg based alloys decreases by introduction of smaller Al atoms. Superposition of these results produces a positive deviation from Vegard's law for Al based alloys and the negative deviation from Vegard's law for Mg based alloys, which results in a significant jump of the lattice parameter at the equiatomic composition [11]. However, it is inappropriate to talk about the lattice parameter for the system formed by the FCC metal, Al, with a single lattice parameter, $a$ – the lattice constant, and the HCP metal, Mg, with two parameters, the lattice constant, $a$ and $c/a$ ratio. In this case description of the atomic volume behavior as a function of composition (the deviation from Zen`s law) is more appropriate because it excludes ambiguity imposed by the different crystal structure of the alloy components. The results of calculations are shown in Fig 3b.

### 3.4. Al-Mn.

There is very limited solubility (~ 3.5 at.%, at 658 ºC) of manganese in aluminum [18, 37], Fig. 4a. The lattice parameters for the quenched $\beta$-Mn based Al-Mn alloys have been reported at 9.65 at.% and 18.4 at.% Al [37]. Good agreement between the calculations (the *terminal* approximation) of the atomic volume occurs at both ends of the concentration range, Fig. 4b. The negative deviation from Zen`s correlates with the negative heat of mixing observed in disordered Al-Mn alloys [38].

### 3.5. Al-Ti.

According to Refs. [19, 37], the aluminum-based Al-Ti solid solution is very restricted., Fig. 5a. The maximum solubility of titanium in aluminum is about 0.2 at.% and is not considered in present study. Maximum solubility of aluminum in titanium, quenched from 1200 ºC, is about 42 at.% [37]. The calculated, within the *terminal* approximation, atomic volume of Ti based solid solution is in a fair agreement with experimental data, Fig. 5b, although above ~ 25 at.% of Al calculations overestimate the observed atomic volume. The significant negative deviation from Zen`s law for Ti based solid solution correlates with the negative heat of mixing measures for Al-Ti solid solutions at room temperature [38].

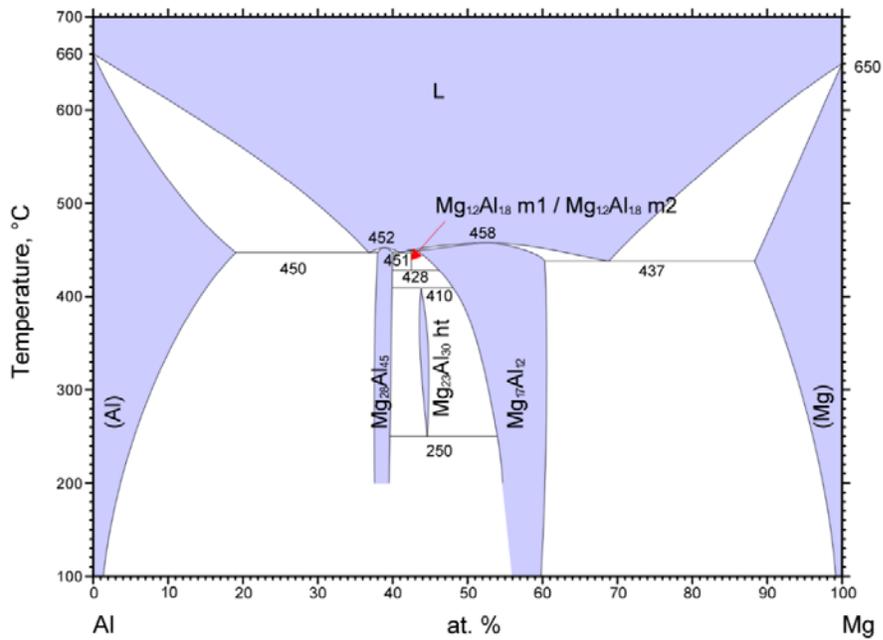

Fig 3a. Al-Mg phase diagram [17]. This plot is taken from ASM Alloy phase Diagram Database.

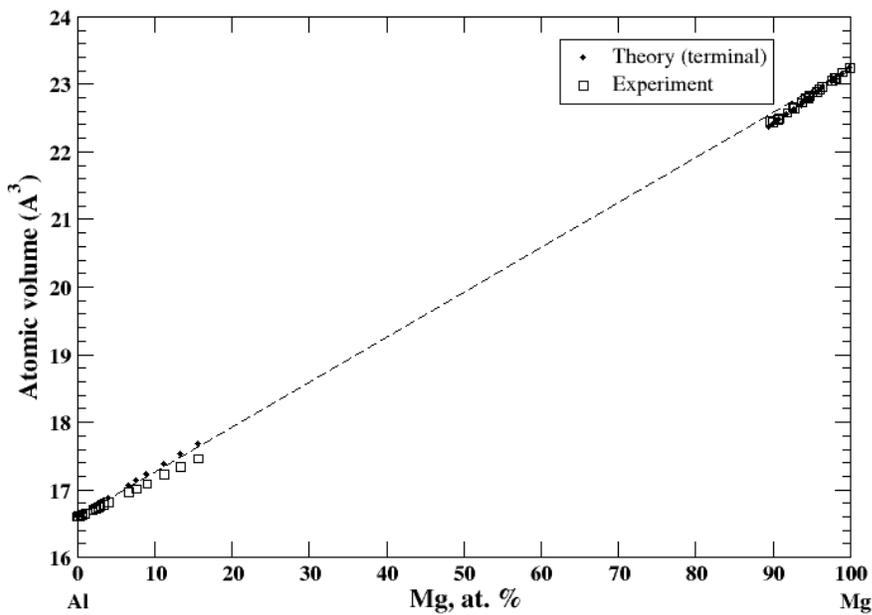

Fig 3b. Atomic volume vs. concentration for Al-Mg alloy system. The experimental data are from Ref. [37], pp. 367 and 728.

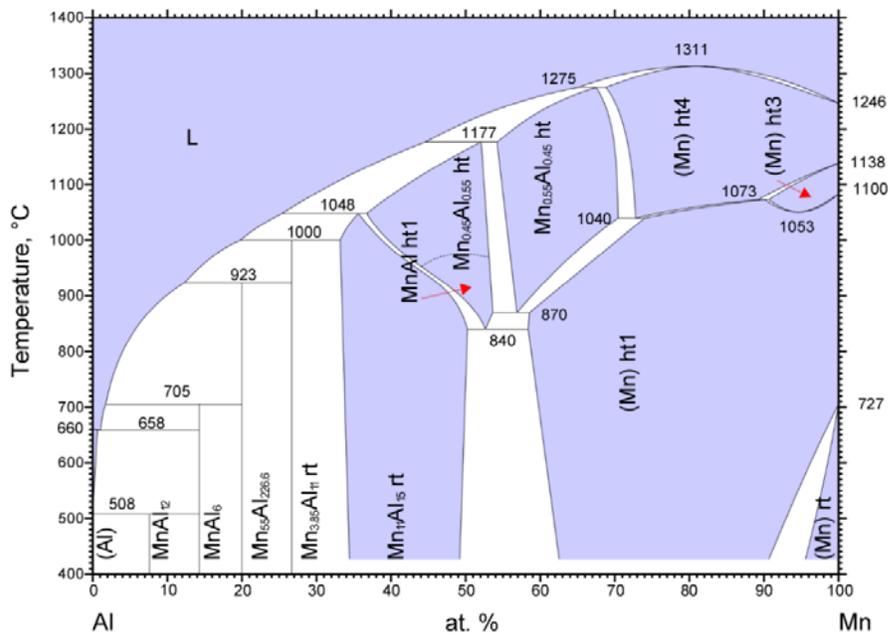

Fig 4a. Al-Mn phase diagram [18]. This plot is taken from ASM Alloy phase Diagram Database.

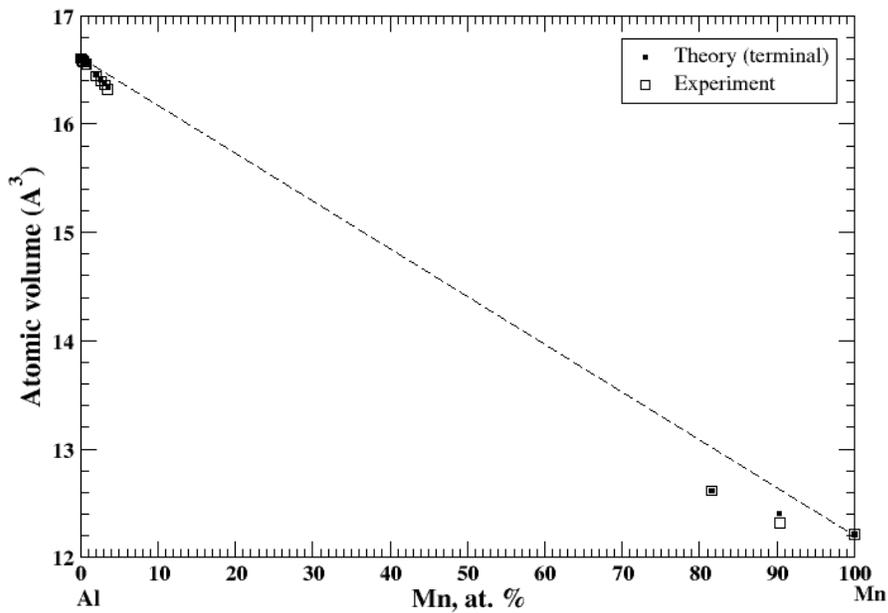

Fig 4b. Atomic volume vs. concentration for Al-Mn alloy system. The experimental data are from Ref. [37], pp. 373 and 374.

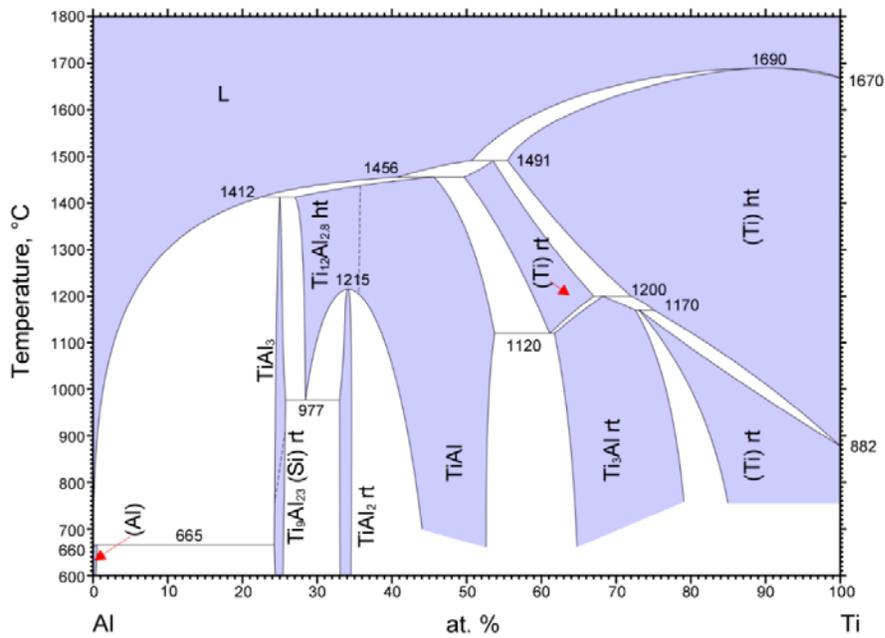

Fig 5a. Al-Ti phase diagram [19]. This plot is taken from ASM Alloy phase Diagram Database.

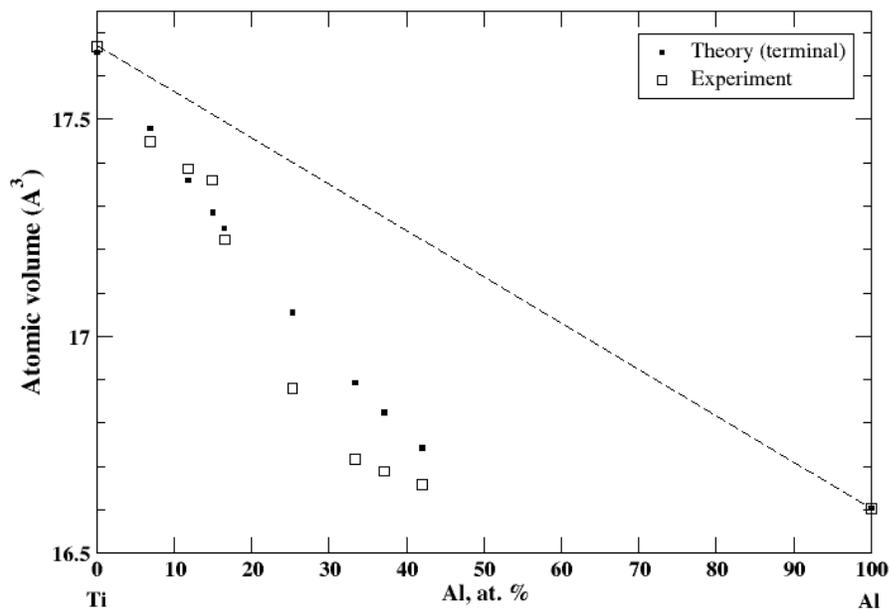

Fig 5b. Atomic volume vs. concentration for Al-Ti alloy system. The experimental data are from Ref. [37], p. 386.

### 3.6. Al-Zn.

According to Refs. [20, 37], there is a slight solid solubility of aluminum in zinc and extensive solubility of zinc in aluminum extending to ~ 65 at.% Zn at 381 ºC, Fig. 6a. For Al- based Al-Zn solid solutions the lattice spacing has been measured for alloys containing up to 35 at.% Zn [37]. For Zn-based Al-Zn solid solutions the lattice spacing has been measured for alloys containing up to ~ 1.7 at.% of Al. The calculated atomic volume of Al-Ti solid solution in the *terminal* approximation is in a good agreement with experimental data, Fig. 6b. The positive deviation from Zen`s law is described for Al-based Al-Zn solid solution in a perfect accord with experimental data. For Zn-based Al-Zn solid solution, the present calculations show a negative deviation from Zen`s law, which again is an accord with experimental data available up to ~ 2 at.% of Al. Al and Zn form a eutectic phase diagram with the small but positive heat of formation for solid Al-Zn alloys [38]. This positive heat of formation correlates with the positive deviation from Zen`s law for the extended range of Al-based Al-Zn solid solutions.

### 3.7. Ag-Cu.

According to Refs. [21, 37], Ag and Cu form the eutectic type phase diagram with restricted terminal solid solutions. The maximum solubility of silver in copper is about 5 at.% and maximum solubility copper in silver is about 14 at.% at the eutectic temperature of 779 ºC (Fig. 7a.). For Cu-based alloys, calculations of the atomic volume, performed within both *continuum* and *terminal* approximation, give an excellent agreement with experimental data measured up to 3.7 at.% of Ag at 770 ºC [37], Fig 7.b. For Ag based alloys, calculations of the atomic volume, performed within the *terminal* approximation, give an excellent agreement with experimental data measured up to 12.1 at % of Cu at 770 ºC [37], Fig 7b. The *continuum* approximation slightly overestimates the atomic volume of Ag-based alloys. The small positive deviation from Zen`s law correlated with a small but positive heat of formation observed in Cu-Ag alloys [38].

### 3.8. Cu-Au.

According to Refs. [22, 37], copper and gold form a continuous solid solution at high temperatures, Fig 8a. The calculated, within *continuum* approximation, atomic volume of Cu-Au solid solution is in a good agreement with experimental data, Fig. 8b, reproducing a slight positive deviation from Zen`s law. The measured at high temperatures, ~ 500 ºC – 700 ºC, heat of formation of continuous solid Cu-Au solutions is negative [38] and the positive deviations from Zen`s law is stipulated exclusively by the size effect.

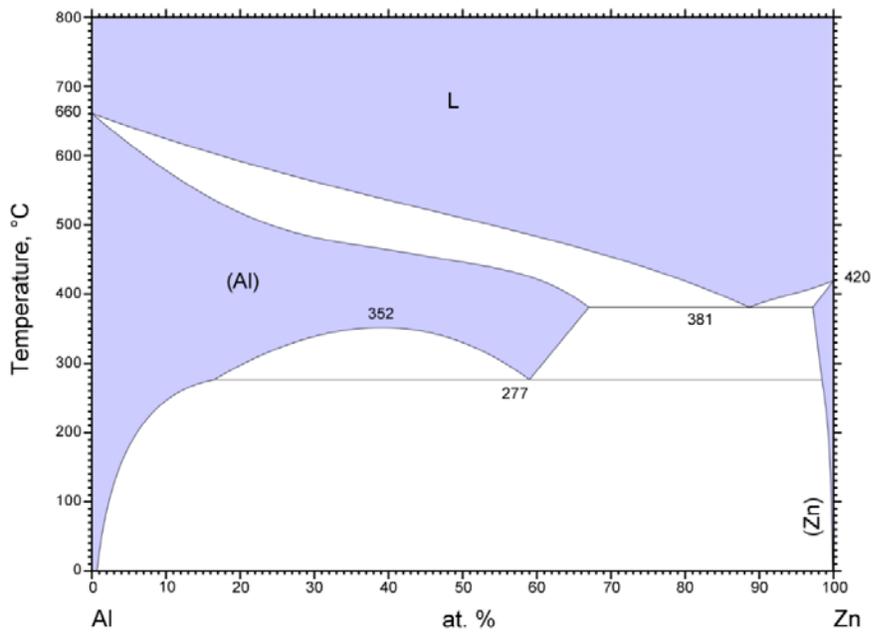

Fig 6a. Al-Zn phase diagram [20]. This plot is taken from ASM Alloy phase Diagram Database.

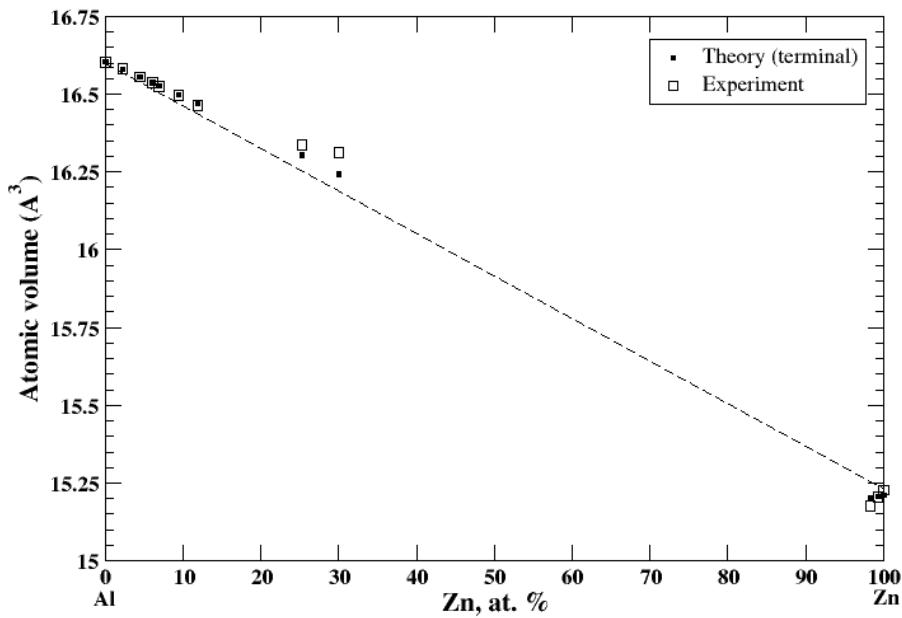

Fig 6b. Atomic volume vs. concentration for Al-Zn alloy system. The experimental data are from Ref. [37], pp. 389 and 886.

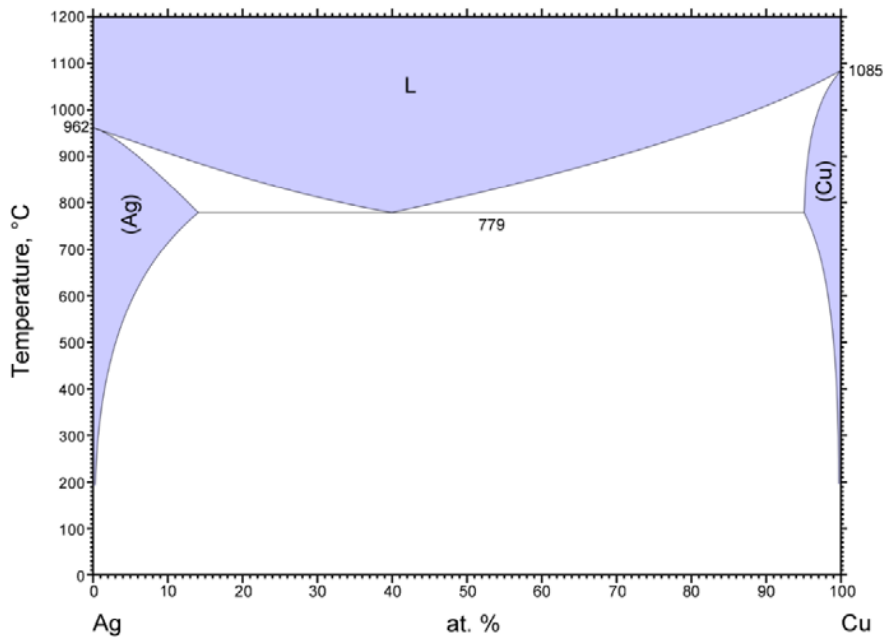

Fig 7a. Ag-Cu phase diagram [21]. This plot is taken from ASM Alloy phase Diagram Database.

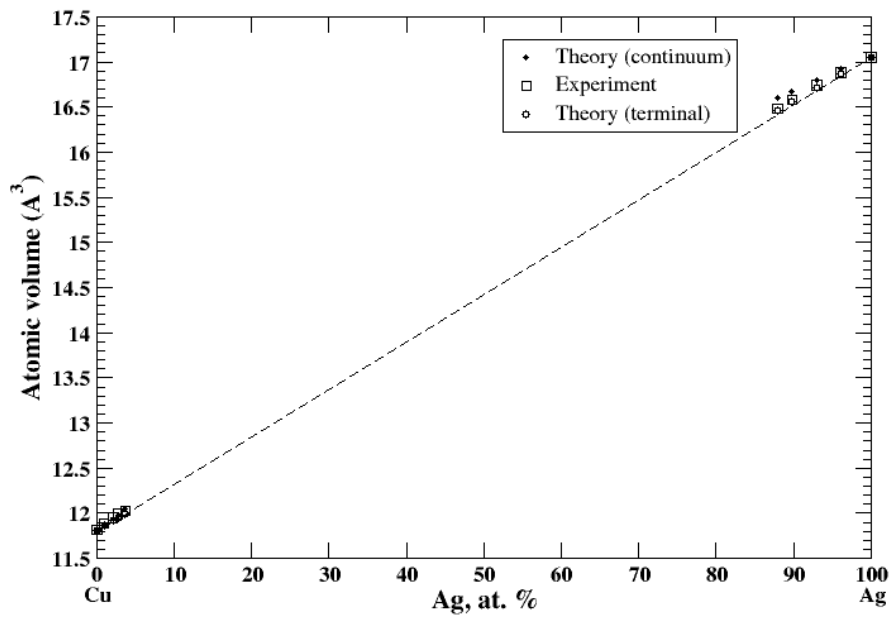

Fig 7b. Atomic volume vs. concentration for Ag-Cu alloy system. The experimental data are from Ref. [37], pp. 279 and 602.

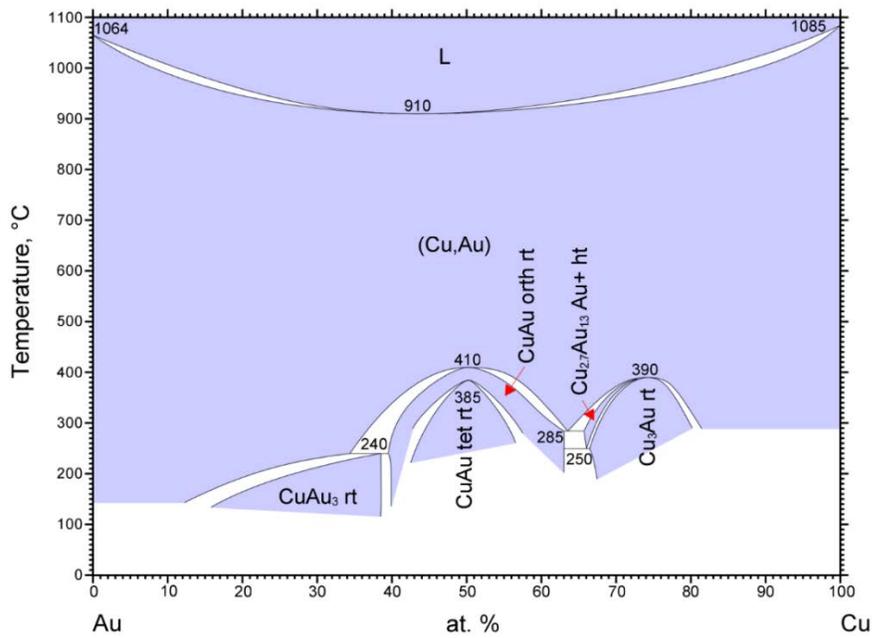

Fig 8a. Au-Cu phase diagram [22]. This plot is taken from ASM Alloy phase Diagram Database.

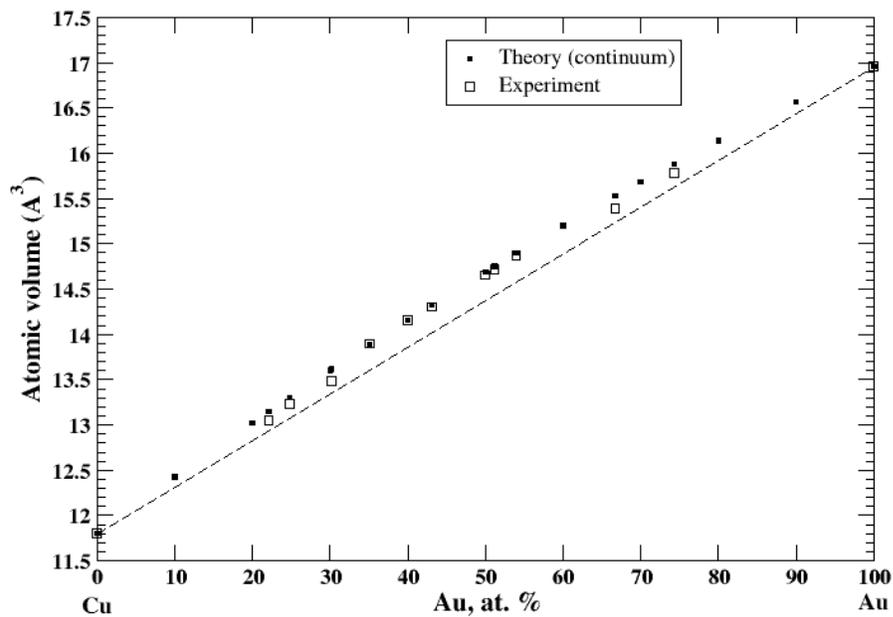

Fig 8b. Atomic volume vs. concentration for Au-Cu alloy system. The experimental data are from Ref. [37], pp. 411 and 601.

### 3.9. Cu-Fe.

According to Ref. [23], the solubility limit of copper in $\alpha$-iron is small, ~ 1.9 at.%. at eutectoid temperature 840 ºC, the solubility of $\alpha$-iron in copper at the same temperature is also small, 1.3 at %, Fig. 9a. At peritectic temperature, 1090 ºC, about 7.5 % of copper can be dissolved in $\gamma$-iron, and about 4.6 at.% of $\gamma$-iron can be dissolved in copper, Fig 9a. The experimental data of the lattice constant for Cu-based alloys is reported up to 2.7 at. of Fe, and for Fe based alloys the lattice constant is reported only up to 0.66 at.% of Cu [37]. The atomic volume, calculated within the *terminal* approximation, reproduces experimental measurements, and shows a positive deviation from Zen`s law, Fig 9b. The very small solubility limits in Cu-Fe solid solutions correlate with a significant positive heat of formation in this system [38].

### 3.10. Cu-Ni.

According to Refs. [24, 37], copper and nickel form a continuous solid solution, Fig. 10a. The calculated, within the *continuum* approximation, atomic volume of Cu-Ni solid solution is in a good agreement with experimental data, Fig. 10b, reproducing a slight negative deviation from Zen`s law. The heat of formation of the Cu-Ni solid solution is moderate positive within the composition range, [38], signaling that the entropy factor plays a decisive role in formation of a continuous solid solution above 365 ºC [38].

### 3.11. Cu-Zn.

According to Refs. [25, 37], Cu-Zn system contains six intermediate phases over a composition range, Fig 11a. Two of them, $\alpha$ and $\eta$, represent solid solutions based on Cu-rich and Zn-rich part of the Cu-Zn phase diagram, respectively. The maximum solubility of Zn- in Cu-based solid solution is about 37 at.% [37]. The maximum solubility of Cu in Zn-based solid solution is about 3 at.% [37]. The atomic volume, calculated within the *terminal* approximation, for both Cu-based and Zn-based solid solutions reproduces experimental measurements and shows a significant negative deviation from Zen`s law, Fig 11b, which is in accord with a significant negative heat of formation of disordered Cu-Zn alloys [38].

### 3.12. Co-Fe.

According to Refs. [26, 37], cobalt and iron form a continuous solid solution at high temperatures, Fig 12a. The calculated, within the *continuum* approximation, atomic volume of Co-Fe solid solution is in a good agreement with experimental data measured at 575 ºC for FCC and BCC solid solutions [37], Fig. 12b, reproducing a positive deviation from Zen`s law. The heat of

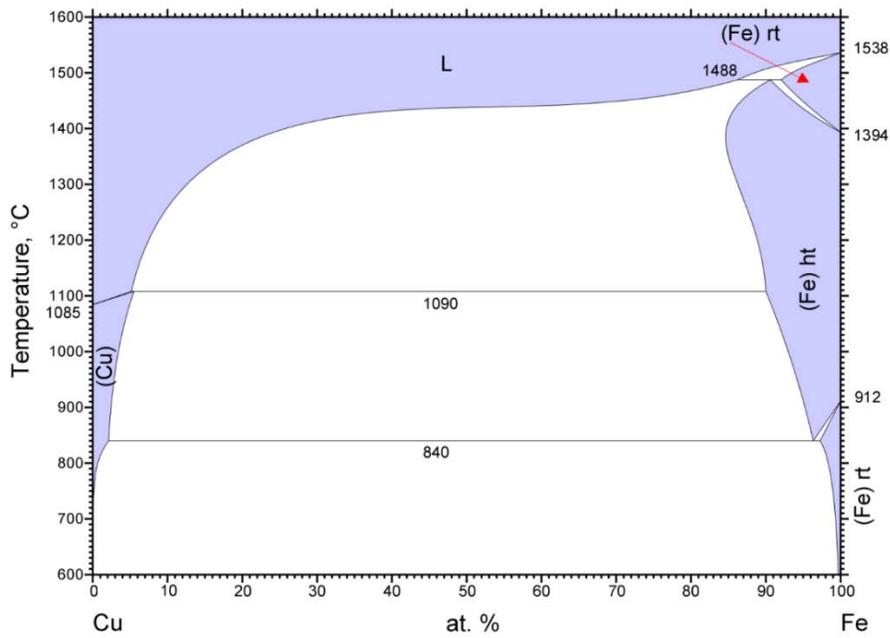

Fig 9a. Cu-Fe phase diagram [23]. This plot is taken from ASM Alloy phase Diagram Database.

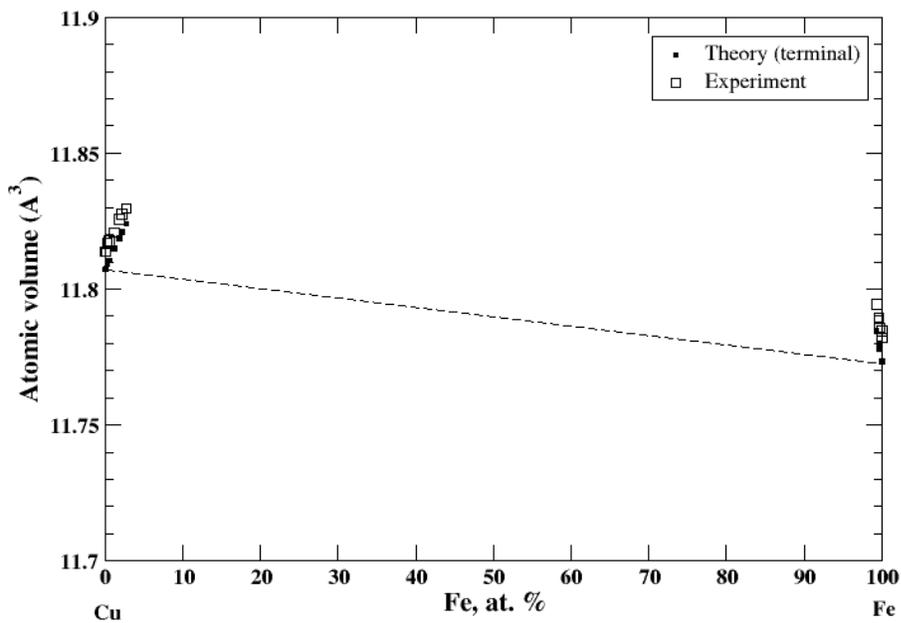

Fig 9b. Atomic volume vs. concentration for Cu-Fe alloy system. The experimental data are from Ref. [37], pp. 571 and 632.

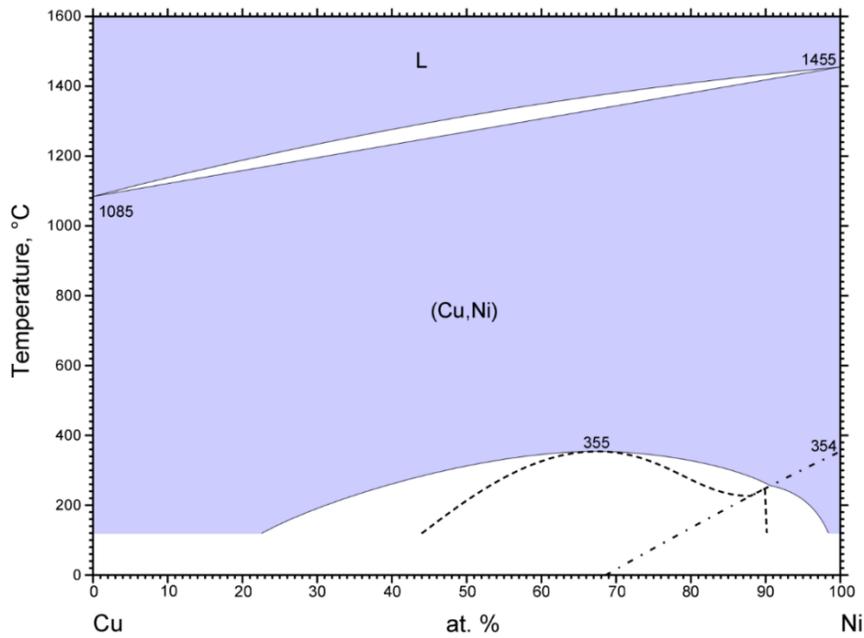

Fig 10a. Cu-Ni phase diagram [24]. This plot is taken from ASM Alloy phase Diagram Database.

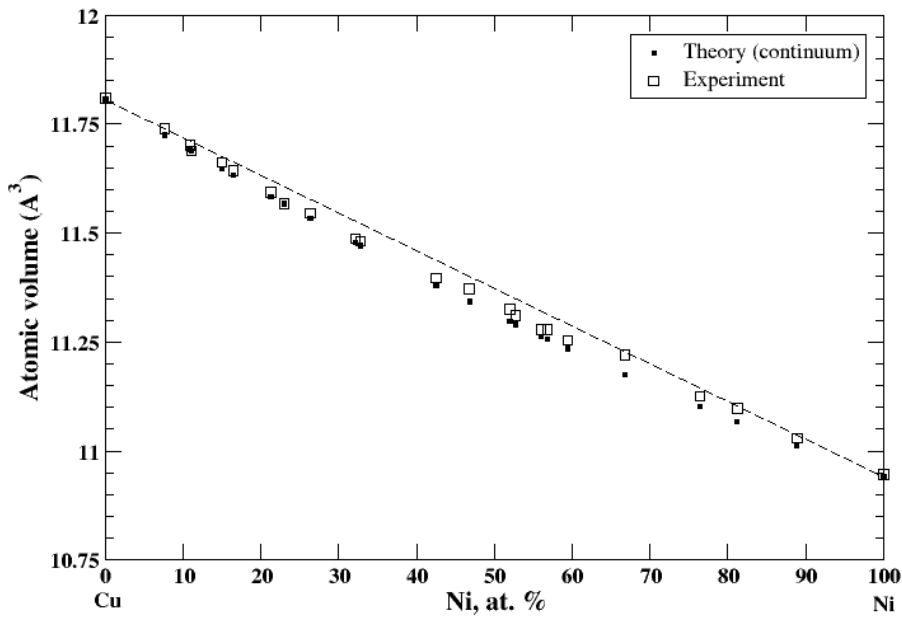

Fig 10b. Atomic volume vs. concentration for Cu-Ni alloy system. The experimental data are from Ref. [37], pp. 592 and 601.

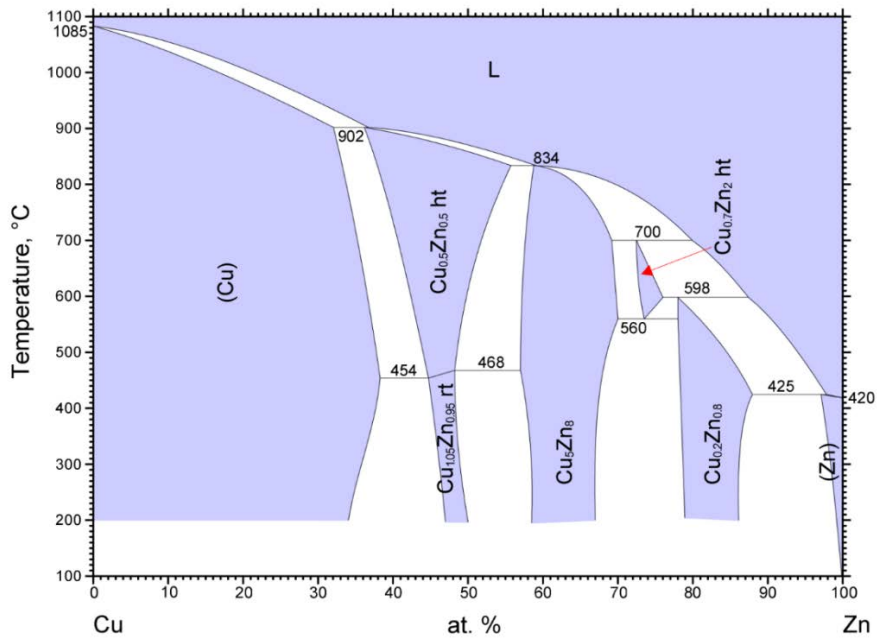

Fig 11a. Cu-Zn phase diagram [25]. This plot is taken from ASM Alloy phase Diagram Database.

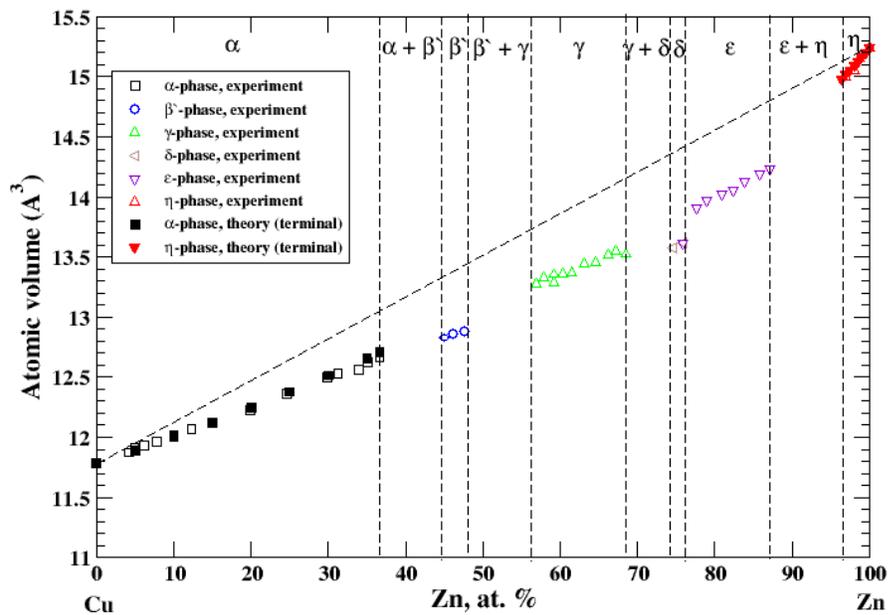

Fig 11b. Atomic volume vs. concentration for Cu-Zn alloy system. The experimental data are from Ref. [37], pp. 620 and 622.

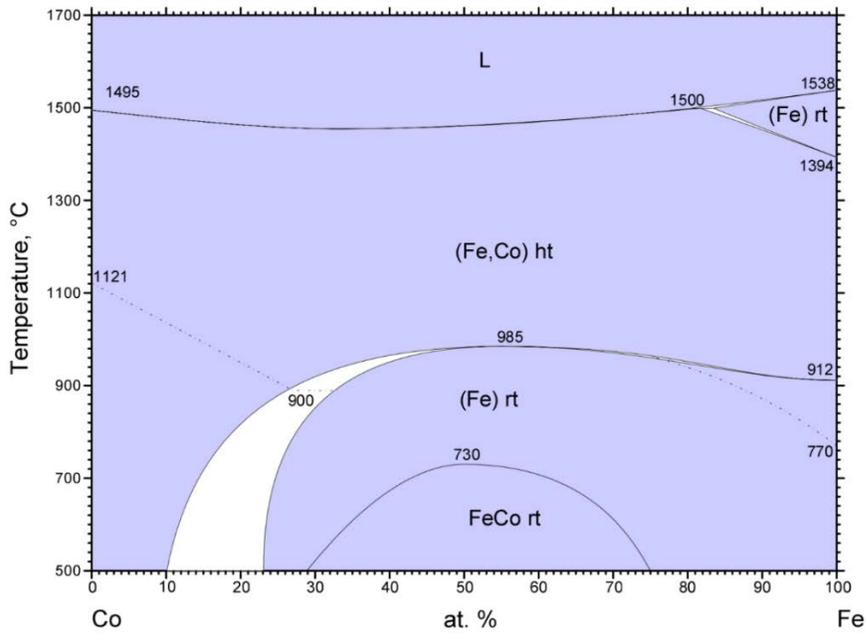

Fig 12a. Fe-Co phase diagram [26]. This plot is taken from ASM Alloy phase Diagram Database.

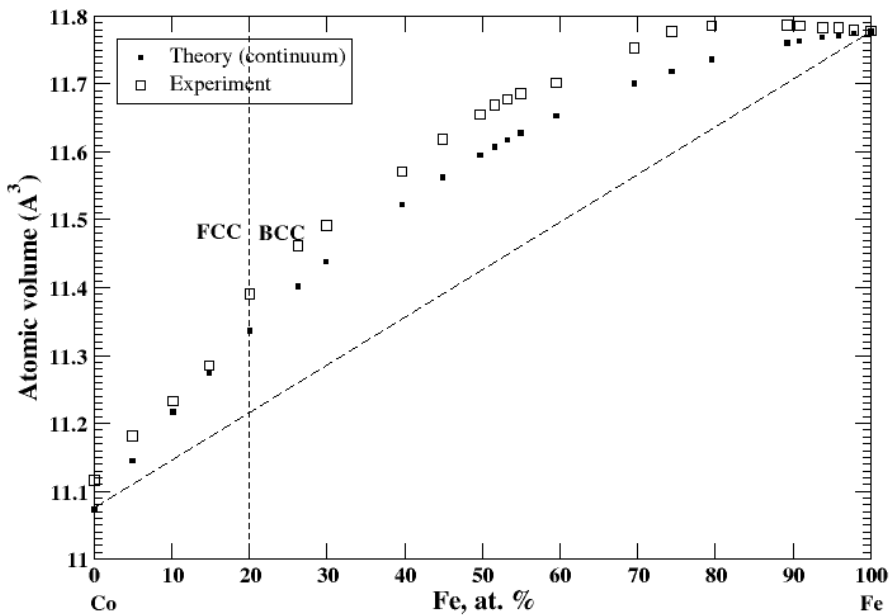

Fig 12b. Atomic volume vs. concentration for Fe-Co alloy system. The experimental data are from Ref. [37], pp. 505 and 634.

formation of liquid Co-Fe alloys at 1590 ºC is a small negative value [38] reflecting a continuous solid solution at high temperature.

### 3.13. Fe-Cr.

According to Refs. [27, 37], iron and chromium form a continuous solid solution at elevated temperatures, Fig 13a. The calculated, within the *continuum* approximation, atomic volume of Fe-Cr solid solution shows a strong positive deviation from Zen`s law which is an accord with experimental data up to ~ 12 at % of Cr, Fig 13b. However, between ~ 12 and 17 at.% of Cr the experimental atomic volume remains almost unchanged, then slightly increases between 17 and 25 at.% of Cr, then drops to its value at ~ 19 at.% of Cr, and then gradually increases within the remaining compositional range, [37]. Fig. 13b also shows calculated, within the *terminal* approximation, volume of Cr based solid solution in the compositional range, 30 at.% -100 at.% of Cr, which are in an excellent agreement with experimental data [37] (and Zen`s law). The heat of formation of Fe-Cr solid solution, measured at 1327 ºC [38], is positive indicating that the entropy factor is responsible for formation of a continuous solid solution at elevated temperatures.

### 3.14. Fe-V.

According to Refs. [28, 37], iron and vanadium form a continuous solid solution at elevated temperatures, Fig 14a. The calculated, within the *continuum* approximation, atomic volume of Fe-V solid solution is in a good agreement with experimental data measured above 1252 ºC [37], Fig. 14b, reproducing a significant negative deviation from Zen`s law. According to [38], the heat of formation of Fe-V solid solution, measured at 1327 ºC, is positive up to ~ 52 at.% of V and slightly negative in the remaining part of the composition range.

### 3.15. Ag-Au.

According to Refs. [29, 37], silver and gold form a continuous solid solution, Fig 15a. The calculated atomic volume in the *continuum* approximation shows a significant negative deviation from Zen`s like one observed experimentally [37], Fig 15b. The heat of formation of the solid solution, measured at 527 ºC [38] is significantly negative.

### 3.16. Ag-Mg.

According to Refs. [30, 37], magnesium dissolved in silver beyond 25 at.% (the maximum solubility of Mg in Ag is ~ 29 at.% at eutectic temperature of 759 ºC, Fig 16a. The maximum solubility of Ag in Mg is significantly smaller, ~ 4 at.% at eutectic temperature 472 ºC, Fig 16a. The atomic volume, calculated within the *terminal* approximation, for both Ag-based and Mg-based

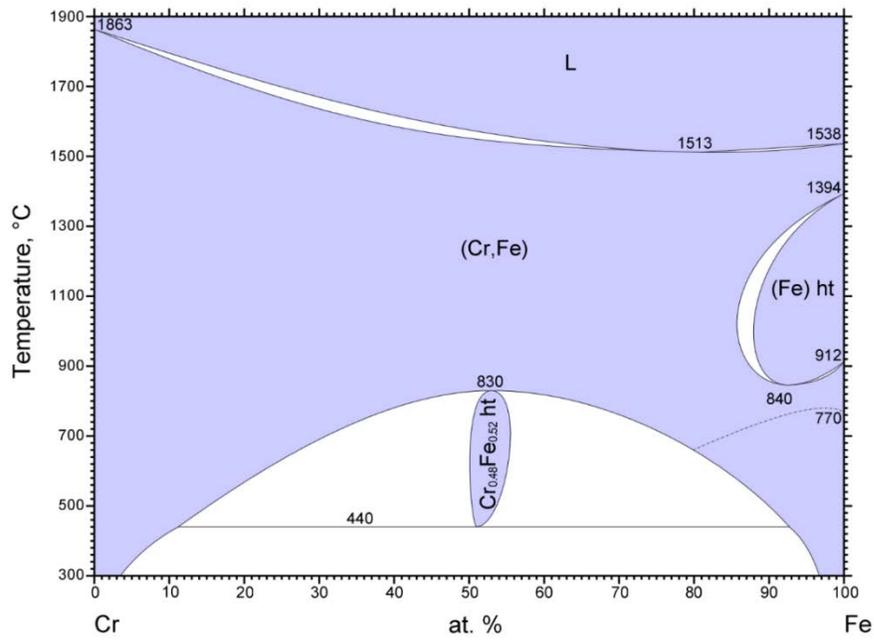

Fig 13a. Fe-Cr phase diagram [27]. This plot is taken from ASM Alloy phase Diagram Database.

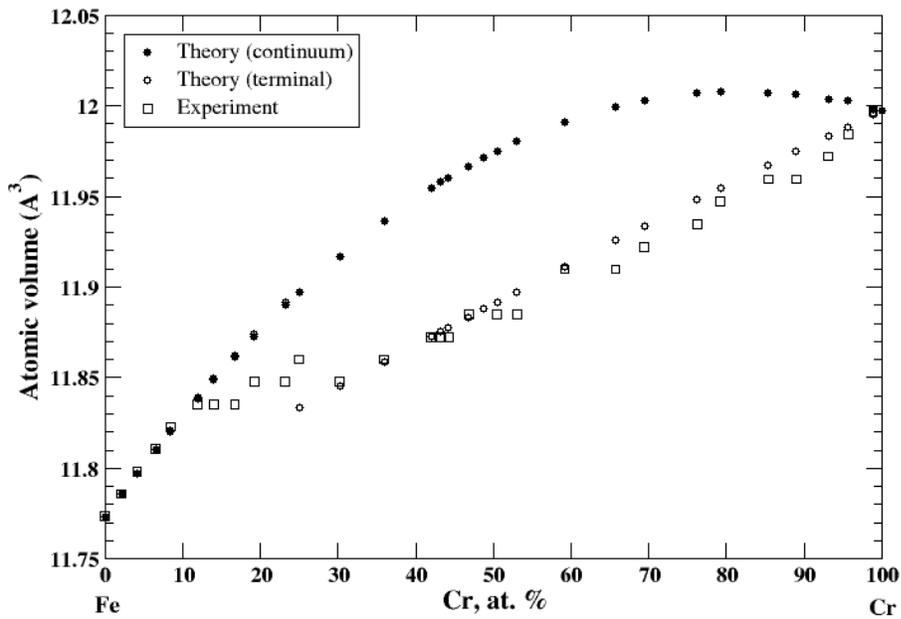

Fig 13b. Atomic volume vs. concentration for Fe-Cr alloy system. The experimental data are from Ref. [37], pp. 533 and 544.

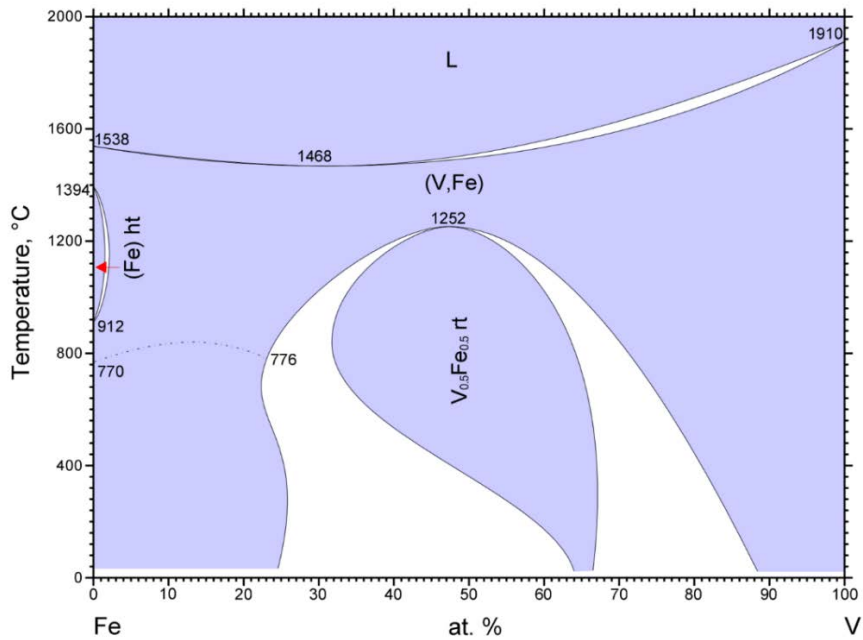

Fig 14a. Fe-V phase diagram [28]. This plot is taken from ASM Alloy phase Diagram Database.

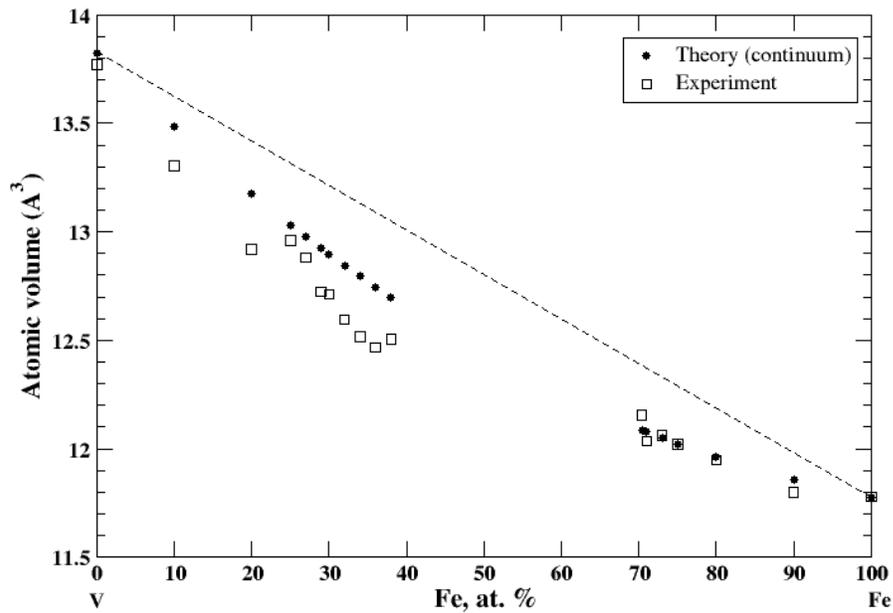

Fig 14b. Atomic volume vs. concentration for Fe-V alloy system. The experimental data are from Ref. [37], pp. 634 and 663.

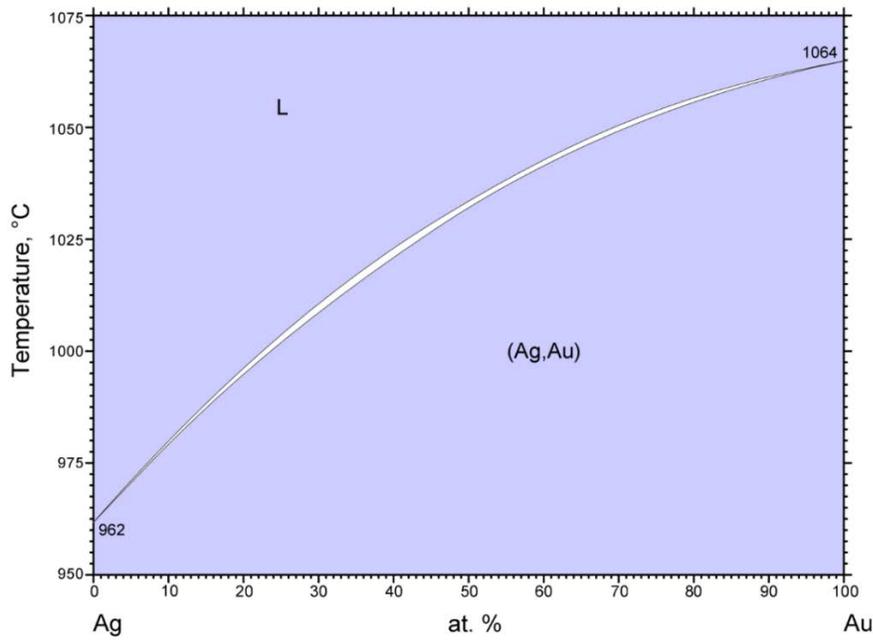

Fig 15a. Ag-Au phase diagram [29]. This plot is taken from ASM Alloy phase Diagram Database.

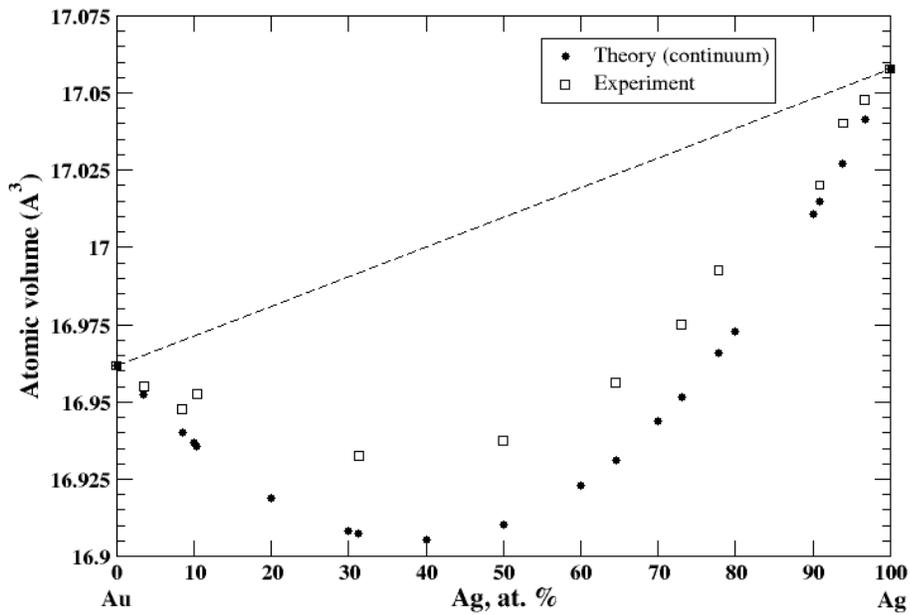

Fig 15b. Atomic volume vs. concentration for Ag-Au alloy system. The experimental data are from Ref. [37], pp. 267 and 289.

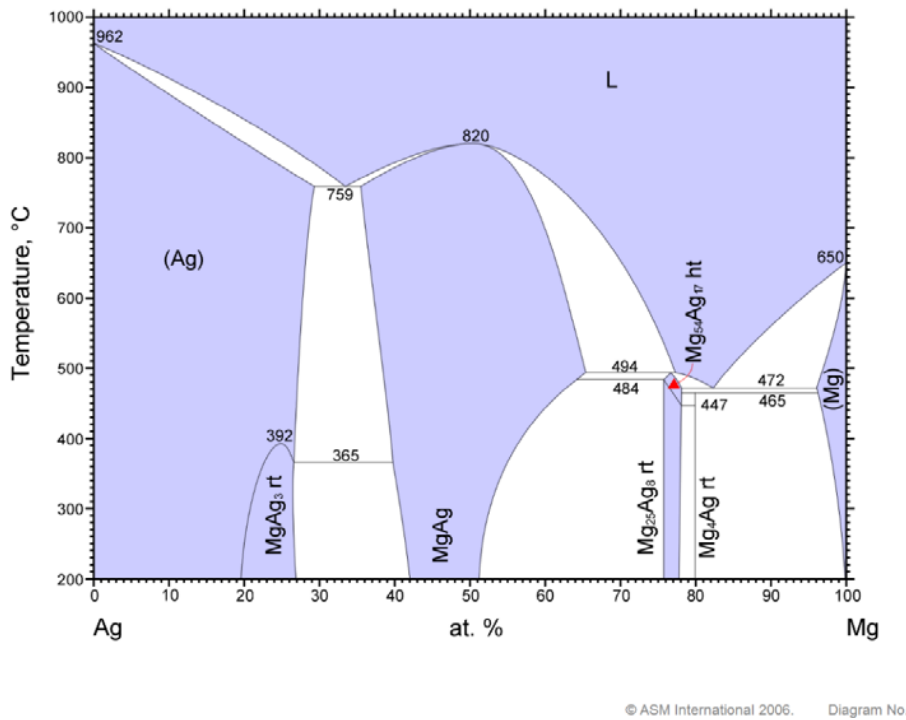

Fig 16a. Ag-Mg phase diagram [30]. This plot is taken from ASM Alloy phase Diagram Database.

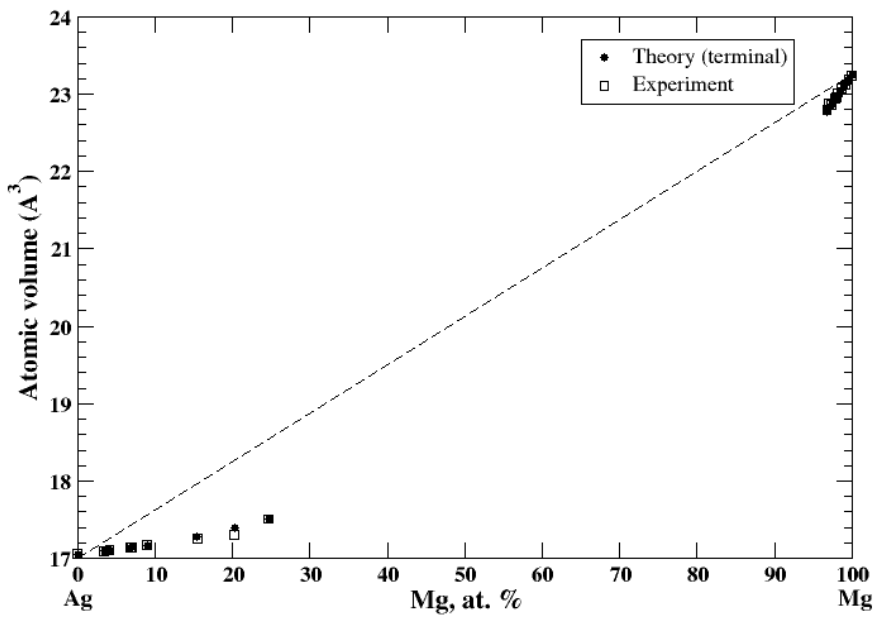

Fig 16b. Atomic volume vs. concentration for Ag-Mg alloy system. The experimental data are from Ref. [37], pp. 292 and 729.

solid solution show the negative deviation from Zen`s law with excellent accord with experimental data [37], Fig 16b. The significant negative heat of formation of Ag based alloys is reported in [38].

### 3.17. Cd-Mg.

According to Refs. [31, 37], cadmium and magnesium form a continuous solid solution at high temperatures, Fig 17a. The calculated, within the *continuum* approximation, atomic volume of Cd-Mg solution show a significant deviation from Zen`s law follow the tendency that was observed experimentally, [37], at 310 ºC, Fig 17b. According to [38], the heat of formation of solid Cd-Mg solution, measures at 270 ºC, is also negative.

### 3.18. Ge-Si.

According to Refs. [32, 37], germanium and silicon form a continuous solid solution, Fig. 18 a. The measured atomic volume, [37], shows a very small negative deviation from Zen`s law which cannot be reproduced either by *terminal*, Fig 18b, or by the *continuum* approximation (not shown). Only if one uses the *real* volumes instead of *apparent* volumes, the experimental results can be reproduced, Fig 18b. Both pure Ge and Si are not metals, contrary to all systems studied here. This is probably a partial explanation of the failure of the *apparent* volume theory.

### 3.19. Nb-Ta.

According to Refs. [33, 37], niobium and tantalum form a continuous solid solution, Fig. 19a. However, the experimental data for the atomic volume are available only for 34.3 at.% and 62.25 at.% of Nb, Fig. 19b. The calculated, within the *continuum* approximation, atomic volume of Nb-Ta solution is also shown on Fig 19b.

### 3.20. Pb-Sn.

According to Refs. [34, 37], lead and tin form a eutectic. The maximum solubility of Sn in Pb is ~ 29 at.% at eutectic temperature 183 ºC, and maximum solubility of Pb in Sn is 1.5 at.% at the same temperature Fig 20a. The atomic volume, calculated within the *terminal* approximation, for both Pb-based and Sn-based solid solution together with experimental is shown in Fig. 20b. For Pb-based alloys the calculated volume follows Zen`s law, in accord with experiment. Slight positive deviation from Zen`s law is observed for Zn-based alloys. The heat of formation of solid Pb-based alloys is positive which is in an accord with eutectic type of the Pb-Sn phase diagram [37].

### 3.21. Ti-Zr.

According to Refs. [35, 37], titan and zirconium form a continuous solid solution, Fig. 21a. The calculated, within the *continuum* approximation, atomic volume of Ti-Zr solution show a slight

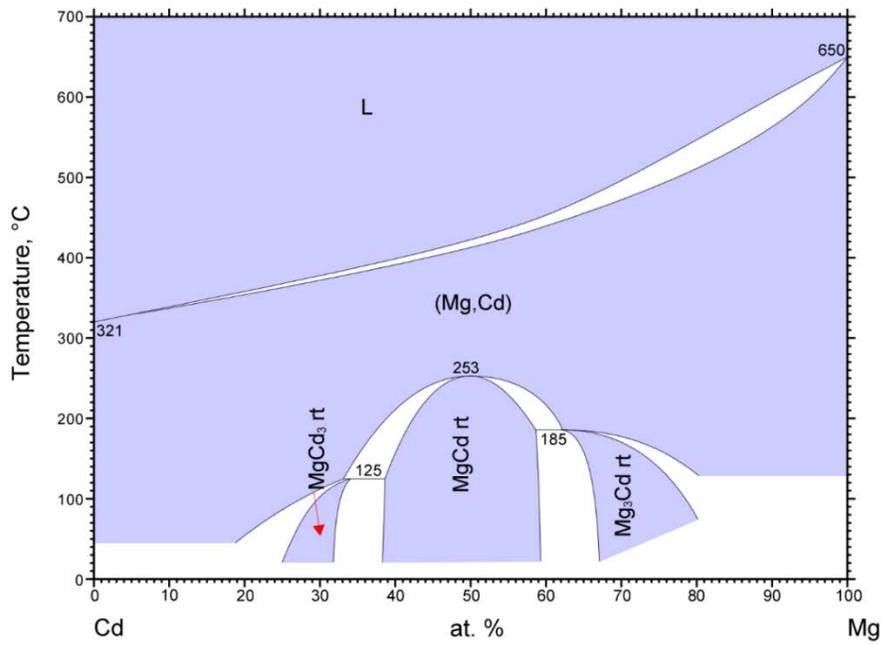

Fig 17a. Cd-Mg phase diagram [31]. This plot is taken from ASM Alloy phase Diagram Database.

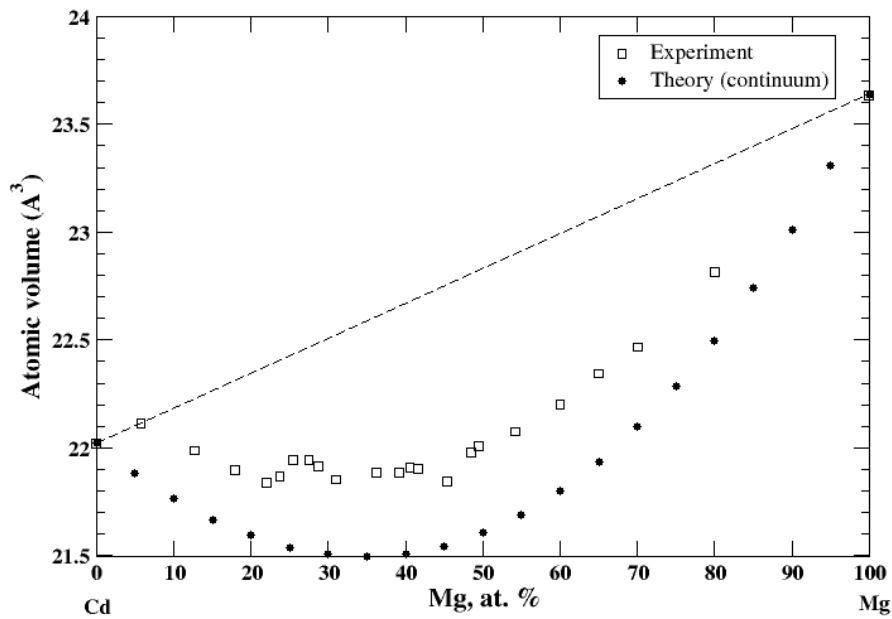

Fig 17b. Atomic volume vs. concentration for Cd-Mg alloy system. The experimental data are from Ref. [37], pp. 49 and 485.

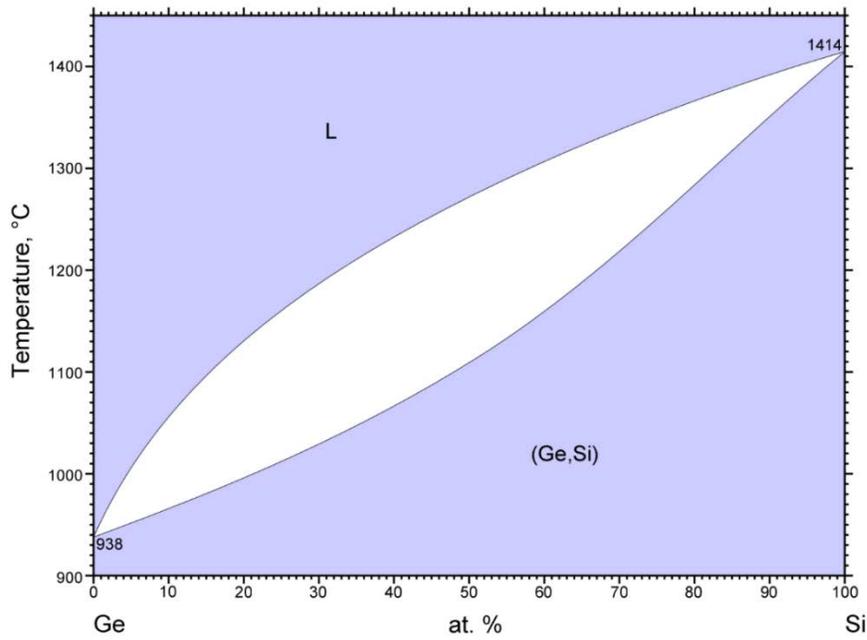

Fig 18a. Ge-Si phase diagram [32]. This plot is taken from ASM Alloy phase Diagram Database.

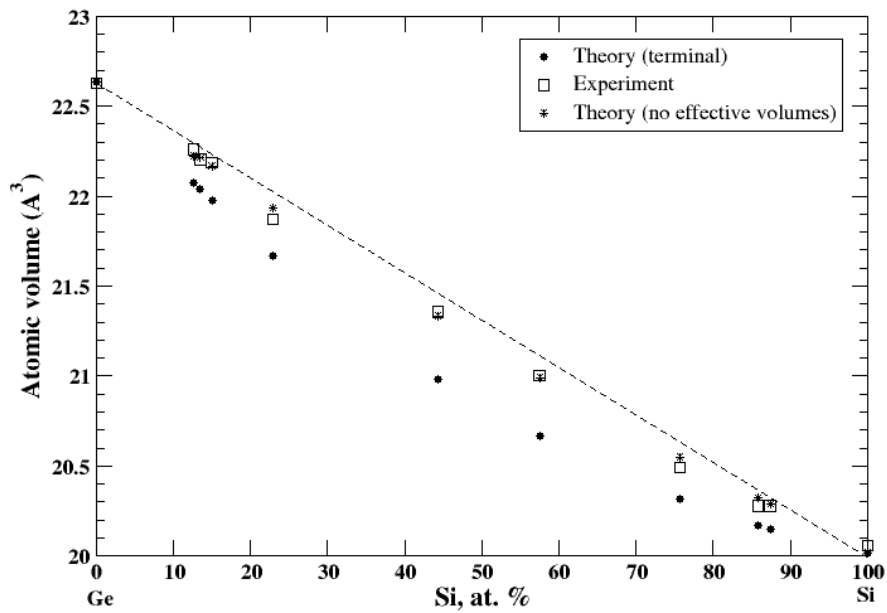

Fig 18b. Atomic volume vs. concentration for Ge-Si alloy system. The experimental data are from Ref. [37], p. 679.

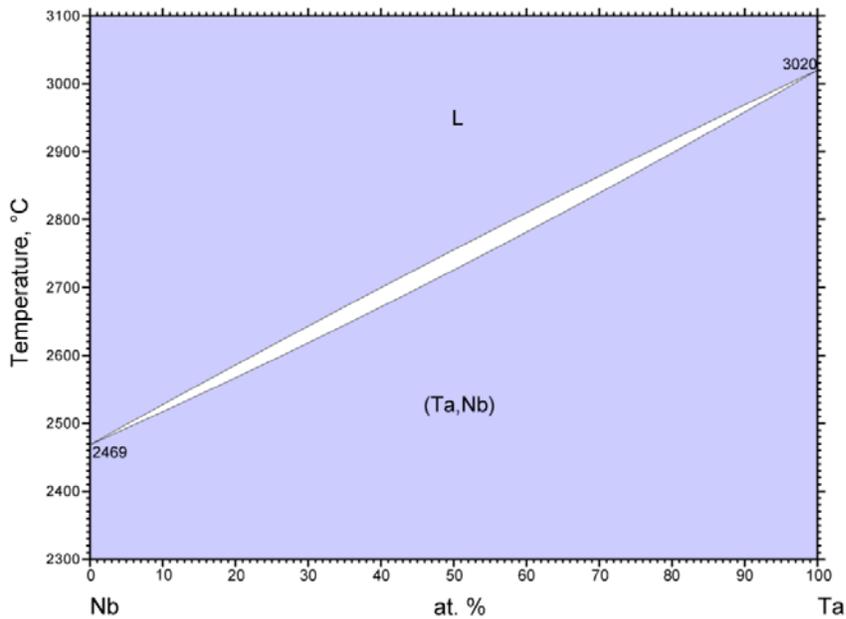

Fig 19a. Nb-Ta phase diagram [33]. This plot is taken from ASM Alloy phase Diagram Database.

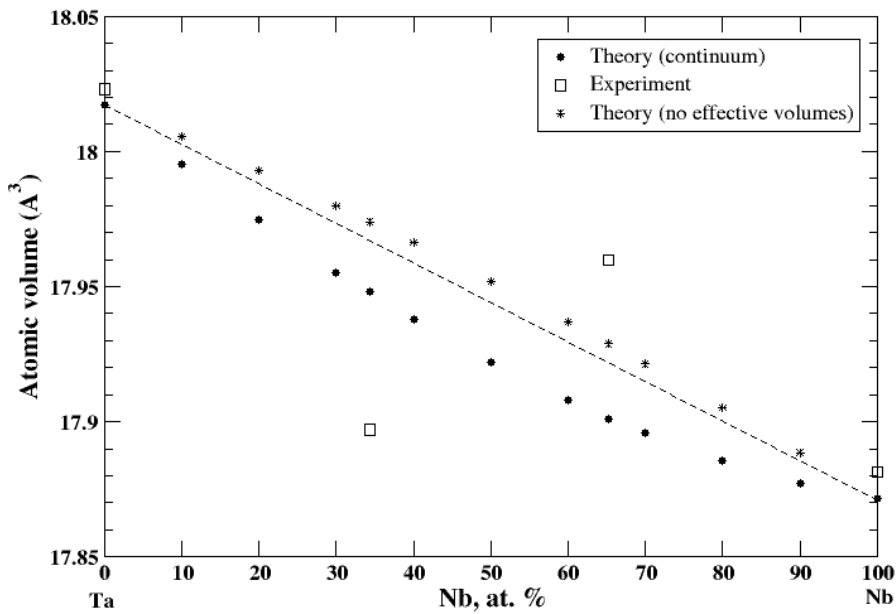

Fig 19b. Atomic volume vs. concentration for Nb-Ta alloy system. The experimental data are from Ref. [37], pp. 757 and 773.

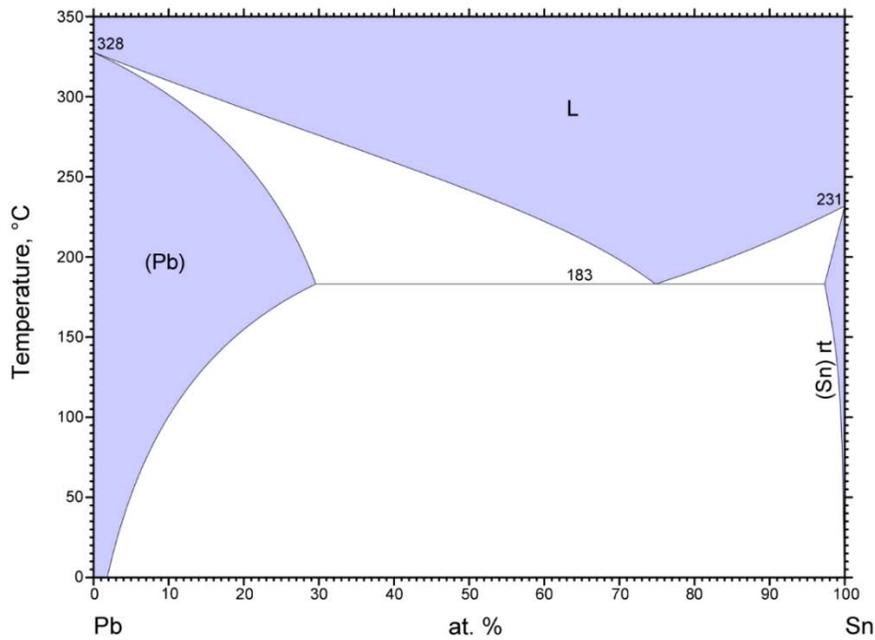

Fig 20a. Pb-Sn phase diagram [34]. This plot is taken from ASM Alloy phase Diagram Database.

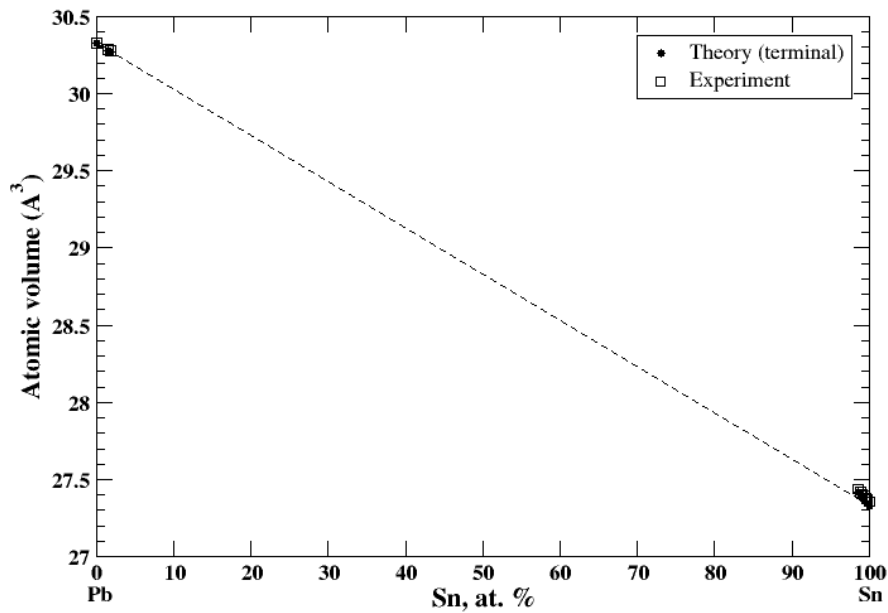

Fig 20b. Atomic volume vs. concentration for Pb-Sn alloy system. The experimental data are from Ref. [37], pp. 818 and 862.

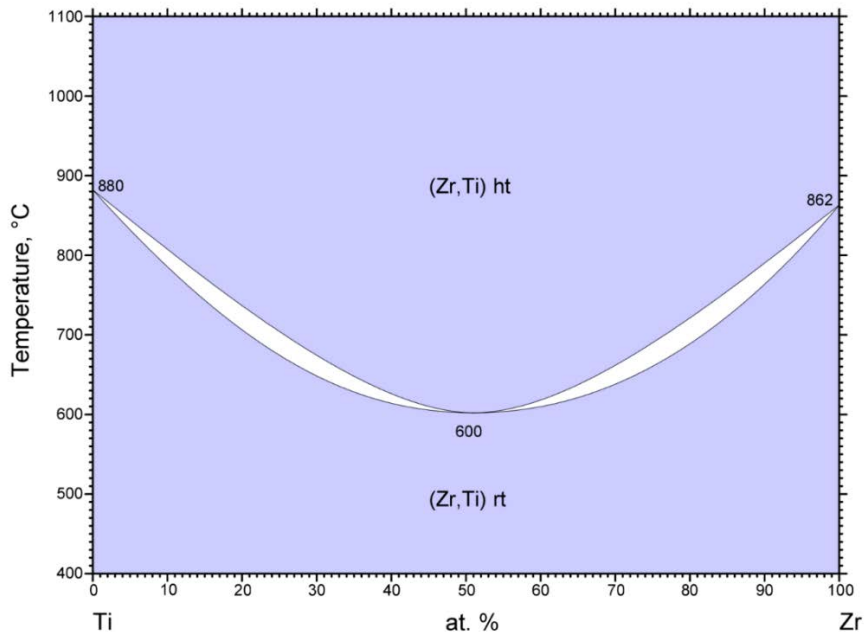

Fig 21a. Ti-Zr phase diagram [35]. This plot is taken from ASM Alloy phase Diagram Database.

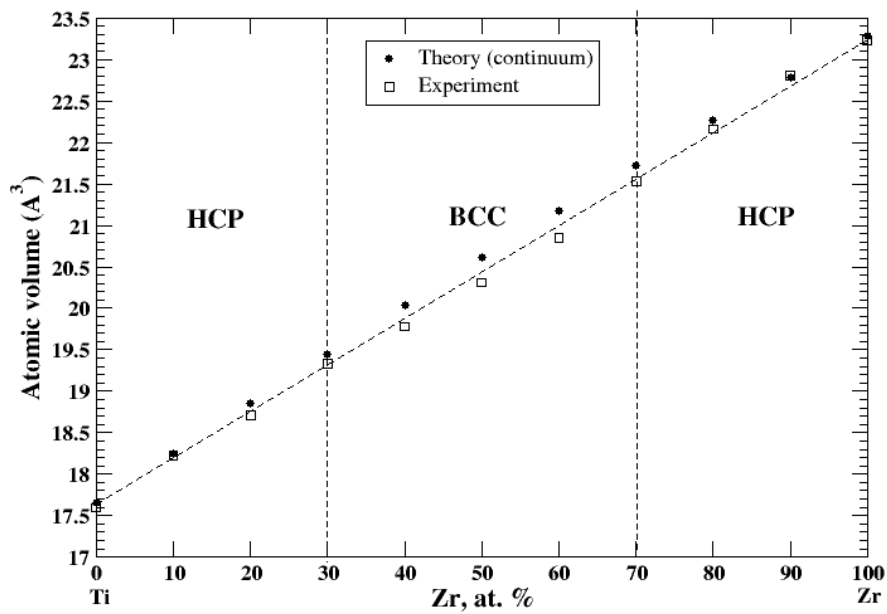

Fig 21b. Atomic volume vs. concentration for Ti-Zr alloy system. The experimental data are from Ref. [37], pp. 873 and 876.

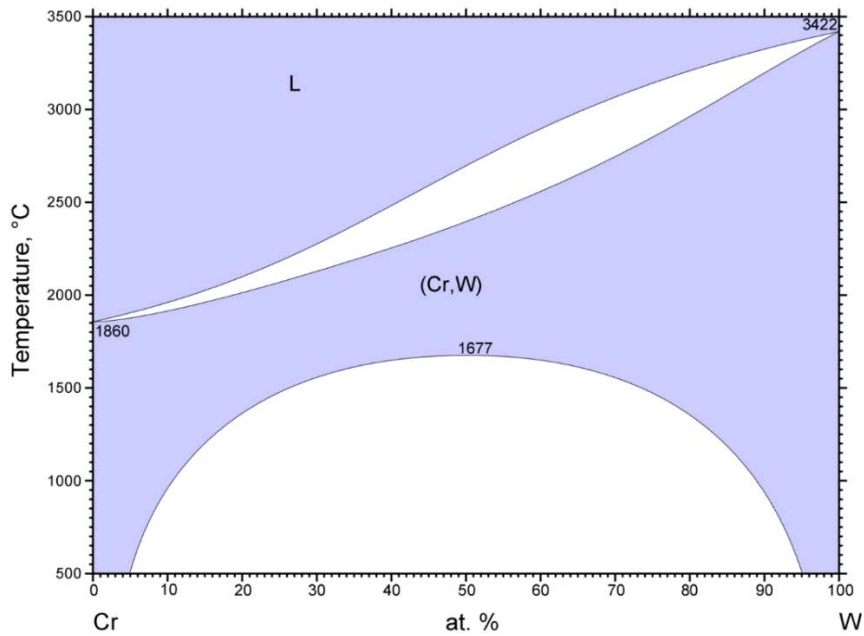

Fig 22a. Cr-W phase diagram [36]. This plot is taken from ASM Alloy phase Diagram Database.

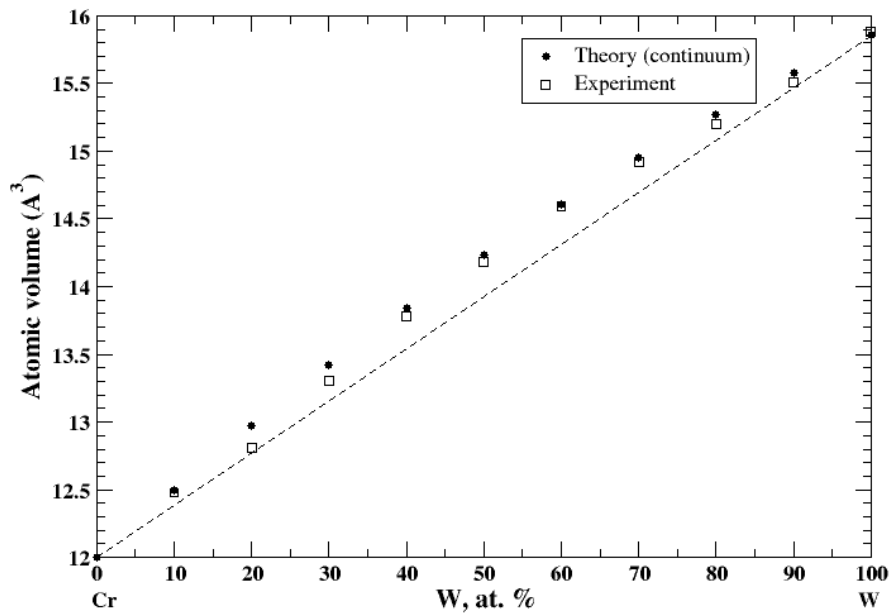

Fig 22b. Atomic volume vs. concentration for Cr-W alloy system. The experimental data are from Ref. [37], pp. 545 and 567.

positive deviation from Zen`s law, although the experimental atomic volume follows Zen`s law, Fig. 21b.

### 3.22. Cr-W.

According to Refs. [36, 37], chromium and tungsten form a continuous solid solution at high temperatures (above 1677 ºC), Fig 22a. The calculated atomic volume of Cr-W solution in the *continuum* approximation shows a positive deviation from Zen`s law which is in an excellent accord with experimental data, Fig. 22b.

## 4. Discussion.

Lubarda [11] studied the validity of Vegard's law in five systems: Al-Cu, Al-Ag, Al-Mg, Cu-Au, and Au-Ag. Four of these alloys are isostructural (FCC) and Al and Mg have FCC and HCP crystal structure, respectively, which causes discontinuity of the lattice constant calculated for Al- and Mg-based terminal solutions where the sign of the deviation from Vegard's law' changes from positive (Al-based) to negative (Mg-based) solid solutions at the equiatomic composition. Even for Al-Cu and Al-Ag isostructural solid solutions, the discontinuity of the lattice constant, calculated for two terminal solutions, is observed at the equiatomic compositions. We show that the study of the deviation from Zen`s law, instead of Vegard's law, removes the problem related to discontinuity of the calculated value (the atomic volume) at the equiatomic composition.

We place twenty-two studied binary alloys in seven different groups: 1. alloys that contain aluminum (Al-Ag, Al-Cu, Al-Mg, Al-Mn, Al-Ti, and Al-Zn); 2. alloys that contain copper (Cu-Ag, Cu-Au, Cu-Fe, Cu-Ni, and Cu-Zn); 3. alloys that contain iron (Fe-Co, Fe-Cr, Fe-V); 4. alloys that contain silver (Ag-Au, Ag-Mg); 5. alloys with both transition metal components (Nb-Ta, Ti-Zr, and Cr-W); 6. alloys which elements belong to the same group of the Periodic Table (Ge-Si, Pb-Sn); and 7) Cd-Mg alloys.

### 4.1. Al-Ag, Al-Cu, Al-Mg, Al-Mn, Al-Ti, and Al-Zn.

These alloys do not form continuous solid solutions. The maximum solubility of aluminum (about 42 at.%) is observed in Al-Ti alloys, although the solubility of titanium in aluminum is very small (0.2 at.%). Al-Ti system show both maximum and minimum mutual solubilities of components for this group of alloys. As we mentioned above, Al-Cu, Al-Ag, and Al-Mg alloys have been discussed in Ref. [11]. The *terminal* approximation works perfectly for all six systems reflecting the correct deviation (the sign) from Zen`s law. For Al-Zn alloys both positive and negative deviations from Zen`s law are reproduced. Al-Mg alloys show the small but negative deviation from Zen`s law, although the positive and negative deviation from Vegard's law is reported [11].

### 4.2. Cu-Ag, Cu-Au, Cu-Fe, Cu-Ni, and Cu-Zn.

The isostructural Cu-Au and Cu-Ni alloys form continuous solid solutions, however Cu-Au alloys show a slight positive deviation from Zen`s law and Cu-Ni alloys show a slight negative deviation from Zen`s law. The *continuum* approximation works perfectly for both these alloys. Cu-Ag and Cu-Fe alloys show limited mutual solubility of the components. The *terminal* approximation works perfectly for both alloys, although the positive deviation from Zen`s law is more pronounced in Cu-Fe alloys due to the very small mutual solubility of the components that have different crystallographic structures, Cu (FCC) and Fe (BCC). Cu (FCC) and Zn (HCP) also have different crystallographic structures, however, there is a significant, about 40 at.%, solubility of Zn in Cu. It is really surprising that *terminal* approximation successfully works within the above-mentioned concentration range, although it is supposed to work well for the limited solubility of the solute on the solvent matrix as occurs for Zn-based solid solution (only approximately 3 at.% of Cu dissolved in Zn).

### 4.3. Fe-Co, Fe-V, and Fe-Cr.

As we already mentioned, iron and vanadium form continuous solid solutions at elevated temperature. Both components, Fe and V, are isostructural (BCC). At ambient temperatures, there is no mutual solubility in the composition region that spans from ~ 40 at.% to ~ 70 at % of Fe. Fe-based alloys, with amount of Fe $\geq$ 70 at.%, are described pretty well within the *continuum* approximations. Two other alloys, Fe-Co and Fe-Cr are formed by magnetic components. Although at room temperature Fe and Co have different structures, BCC and HCP, respectively, iron transforms to FCC structure at elevated temperatures that cause the mutual solubility of the components of Fe-Co alloys within the whole composition range. The *continuum* approximation describes a positive deviation from Zen`s law. Fe and Cr are both isostructural at room temperature, however due to existence of the complex narrow σ-phase in the vicinity of the equiatomic composition, the continuous Fe-Cr solid solutions are formed only above 830 ºC. Both *continuum* and *terminal* approximations work well in the Fe-rich side of the Fe-Cr phase diagram, up to 12 at.% of chromium, describing the strong positive deviation from Zen`s law, however due to unusual behavior of the atomic volume of this system in the composition region between 12 at.% and 25 at.% of Cr, see Ref. [39] for details, both approximations fail in that region. On the remaining part of the Fe-Cr phase diagram, above 25 at % of Cr, the *terminal* approximation (Cr-based alloys) gives excellent agreement with experimental data (and Zen`s law). This is only the case where the *terminal* approximation works better than the *continuum* approximation for alloys with mutual solubility of the components.

### 4.4. Ag-Au and Ag-Mg.

Both Ag and Au are isostructural (FCC) metals and belong to the same subgroup of the Periodic Table. They form continuous Ag-Au solid solutions that are described with the *continuum* approximation (the negative deviation from Zen`s Law). Ag and Mg have different structural modifications, FCC and HCP, respectively. Thus, a limited mutual solubility is observed in Al-Mg alloys. The *terminal* approximation works well in both end points of the Al-Mg phase diagram.

### 4.5. Nb-Ta, Ti-Zr, and Cr-W.

Both Ti and Zr are isostructural (HCP) metals and belong the IV subgroup of the Periodic Table. Cr and W are also isostructural (BCC) and belong to the VI subgroup of the Periodic Table. The composition dependence of the atomic volume is described by the *continuum* approximations well, although the almost perfect Zen`s law behavior is observed (and described) in Ti-Zr alloys and a slight positive deviation is observed (and described) in Cr-W alloys. The situation with Nb-Ta alloys is more complex. Both Nb and Ta metals are isostructural (BCC) and belong to the same V subgroup of the Periodic Table. According to the phase diagram [33], Nb and Ta form continuous solid solutions. However according to Ref. [37], the experimental data for atomic volume is available for two compositions of Nb-Ta alloys only: the negative deviation from Zen`s law is observed at ~ 34 at.% of Nb and the positive deviations from Zen`s law is observed at ~ 62 at.% of Nb. The *continuum* approximation shows the negative deviations from Zen`s law.

### 4.6. Ge-Si and Pb-Sn.

Both Pb and Sn belong to the same 4A group of the Periodic Table and form the eutectic phase diagram. At room temperature the mutual solubilities of the components are negligible small, and the composition dependence of the atomic volume is described well within the *terminal* approximation. Ge and Si also belong the same 4A group of the Periodic Table however, contrary to Pb-Sn system, Ge-Si alloys form continuous solid solutions. As we already mentioned, both *continuum* and the *terminal* approximations could not reproduce the very small negative deviation from Zen`s law, so we speculate that this failure is due to the non-metallic nature of both Si and Ge.

### 4.7. Cd-Mg.

In 1940 Hume-Rothery and Raynor [40] found that the experimental atomic volumes of Mg-Cd solid solutions are smaller than one calculated using the additivity rule (the negative deviation from Zen`s law formulated in 1956, Ref. [3]). Since then, the behavior of Mg-Cd disordered solid solutions become the subject of numerous investigations [41-44]. These works used the pseudopotential method in conjunction with the thermodynamic perturbation theory (Gibbs-Bogoliubov inequality) to calculate the equation of state of the disordered solid and liquid $Mg_xCd_{1-x}$ alloys. The calculated composition dependence of the equilibrium volume of the solid $Mg_xCd_{1-x}$ alloys [44] shows a negative deviation from Zen`s law but not to such an extent as was reported in the experiment, Ref. [40]. The calculations [41-44] have been performed within the local pseudopotential approximations which excluded the charge transfer between alloy components due to the difference of their electronegativity, see Ref. [45] for details. Incorporation of the *apparent* size of solute atom, suggested in Ref. [11], together with modifications suggested in the present study, Eqs. (28-36), allows, for the first time, describe the negative deviation form Zen`s law in Mg-Cd solid alloys

### 5. Conclusion.

We have derived an analytical expression for the atomic volume of the binary alloys at the arbitrary composition for use in the equation of state modeling. We wanted this expression to be robust and predictive even in the absence of experimental data at particular concentration. This paper tests our proposed expression by comparison with experimental data for the binary alloys.

There are numerous papers dedicated to the validity of Vegard's and Zen`s law, e.g. [46-49]. Lubarda [11] introduced an *apparent* size of the solute atom in order to account for the electronic interactions between the outermost quantum shells of the solute and solvent atoms. This idea reflects, to some extent, the electron density rearrangement due to the charge transfer in order to cancel the chemical potential difference due to alloying [46-49]. Jacob *et al.* [49] came to conclusion that both Vegard's and Zen`s laws should be downgraded to an approximation which is valid in specific conditions. We agree with this upshot by describing numerous cases of the deviation of Zen`s law with a satisfactory way to describe (or predict) these deviations which is the primary motivation for this study.

**Acknowledgements:** This work was performed under the auspices of the U.S. Department of Energy by Lawrence Livermore National Laboratory under Contract DE-AC52-07NA27344.